\documentclass[aps,prd,twocolumn,showpacs,nofootinbib]{revtex4}
\usepackage{amsmath}
\usepackage{amssymb}
\usepackage{graphicx}
\usepackage{hyperref}
\usepackage{color}
\usepackage{physics}
\usepackage{hubbleflow}
\usepackage[capitalize]{cleveref}

\hypersetup{
    bookmarks=true,         
    unicode=false,          
    pdftoolbar=true,        
    pdfmenubar=true,        
    pdffitwindow=false,     
    pdfstartview={FitH},    
    pdfnewwindow=true,      
    colorlinks=true,       
    linkcolor=red,          
    citecolor=cyan,        
    filecolor=magenta,      
    urlcolor=blue,           
    linktocpage=true
}

\begin{document}

\title{Slow-Roll Inflation at N3LO}

\author{Pierre Auclair} \email{pierre.auclair@uclouvain.be}
\affiliation{Cosmology, Universe and Relativity at Louvain (CURL),
  Institute of Mathematics and Physics, University of Louvain, 2 Chemin
  du Cyclotron, 1348 Louvain-la-Neuve, Belgium}

\author{Christophe Ringeval} \email{christophe.ringeval@uclouvain.be}
\affiliation{Cosmology, Universe and Relativity at Louvain (CURL),
  Institute of Mathematics and Physics, University of Louvain, 2 Chemin
  du Cyclotron, 1348 Louvain-la-Neuve, Belgium}

\date{\today}

\begin{abstract}
  The next generation of cosmological observations will be sensitive
  to small deviations from a pure power law in the primordial power
  spectrum of the curvature perturbations. In the context of slow-roll
  inflation, these deviations are expected and correspond to the
  so-called running of the spectral index. Their measurement would
  bring as much information as the discovery of deviations from scale
  invariance. However, robust parameter inference requires to
  marginalize over any possible higher order uncertainties, which have
  been, up to now, not fully determined.

  We tackle this issue by deriving the inflationary scalar and tensor
  slow-roll power spectra at next-to-next-to-next to leading order
  (N3LO), fully expanded around an observable pivot wavenumber, for
  all single field inflationary models having minimal and non-minimal
  kinetic terms. Our result therefore encompasses string-inspired
  inflationary models having a varying speed of sound.
\end{abstract}

\maketitle

\section{Introduction}
\label{sec:intro}

The final release of the Planck Legacy Cosmic Microwave Background
(CMB) data has made Cosmic Inflation the most favored scenario of the
early universe~\cite{Starobinsky:1979ty, Starobinsky:1980te,
  Guth:1980zm, Linde:1981mu, Albrecht:1982wi, Linde:1983gd,
  Mukhanov:1981xt, Mukhanov:1982nu, Starobinsky:1982ee, Guth:1982ec,
  Hawking:1982cz, Bardeen:1983qw}. Inflation implies that all forms of
matter and radiation observed today, as well as their large scale
structures, are the outcome of coupled quantum fluctuations of both
the metric and a yet unknown scalar degree of freedom around a quasi
de-Sitter empty universe. Inflation can be realized in multiple ways,
ranging from modified gravity effects, multifield models, to membranes
moving in higher dimensional spacetimes~\cite{DeFelice:2010aj,
  Martin:2013tda, Martin:2013nzq, Baumann:2014nda,
  Vennin:2015vfa}. However, the Planck Legacy data come with strong
constraints on various observables that could have been the signature
of non-linear physics in the early universe. As of today, the CMB
anisotropies show no detectable primordial non-Gaussianities, no
peculiar features, and, primordial isocurvature modes are severely
constrained to be small~\cite{Planck:2018jri, Planck:2019kim}. In
other words, the data favor the simplest vanilla inflationary models
in which a single scalar field slowly rolls down a smooth
potential. Obviously, this does not imply that other physics is not at
work, but, it has to be subdominant. Searching for small effects will
then require us to have at our disposal precise theoretical predictions
for the dominant one; namely, slow-roll inflation. This is the
context of this paper.

Slow-roll inflation is a landscape in itself, it is the natural
outcome of many self-gravitating single field
models~\cite{East:2015ggf, Linde:2017pwt, Chowdhury:2019otk,
  Aurrekoetxea:2019fhr, Joana:2020rxm, Joana:2022uwc} and hundreds of
scenarios have been proposed so far~\cite{Martin:2013tda,
  Martin:2013nzq}. However, there exists a unified treatment to
perturbatively derive the expected scalar and tensor power
spectra. The method has been pioneered in
Ref.~\cite{Starobinsky:1979ty} for the tensor modes and in
Refs.~\cite{Mukhanov:1985rz, Mukhanov:1988jd} for the scalar modes,
then extended to include next-to-leading order corrections in
Refs~\cite{Stewart:1993bc, Liddle:1994dx, Nakamura:1996da,
  Gong:2001he, Hoffman:2000ue, Schwarz:2001vv, Leach:2002ar,
  Schwarz:2004tz}. It has also been tested with other approximation
schemes and applied to various classes of models in
Refs.~\cite{Stewart:2001cd, Martin:2002vn, Habib:2002yi, Habib:2004kc,
  Choe:2004zg, Casadio:2004ru, Easther:2005nh, DiMarco:2005nq,
  Casadio:2005xv, Chen:2006nt, Battefeld:2006sz, Kinney:2007ag,
  Yokoyama:2007uu, Lorenz:2008et, Tzirakis:2008qy, Agarwal:2008ah,
  Chiba:2008rp, Ichikawa:2008iq, Langlois:2008qf, DeFelice:2011bh,
  Martin:2013uma, Jimenez:2013xwa, Karam:2017zno}. The perturbative
expansion uses the so-called Hubble flow functions (also abusively
referred to as the slow-roll parameters when evaluated at a given
time) defined by
\begin{equation}
\eps{i+1}(N) \equiv \dv{\ln |\eps{i}|}{N}, \qquad
\eps{1}(N) \equiv -\dv{\ln H}{N}\,.
\label{eq:hubbleflow}
\end{equation}
Here $H=\dot{a}(t)/a(t)$ stands for the Hubble parameter, $N = \ln(a)$
is the e-fold number and $a(t)$ the
Friedmann--Lema\^{\i}tre--Robertson--Walker (FLRW) scale factor during
inflation. For a quasi-de Sitter expansion, $H(N)$ is almost constant
and the Hubble flow functions quantitatively encode how much inflation
deviates from de Sitter. For any single field model, their functional
shape can be (perturbatively) calculated from the field's
potential~\cite{Liddle:1994dx} while they are exactly determined by
the equation-of-state parameter in fluid representations of
single-field inflation~\cite{Mukhanov:2013tua,
  Martin:2016iqo}. Without yet entering into details, $H$ (leading
order) fixes the amplitude of the fluctuations, $\eps{1}$ makes some
corrections to that amplitude, sets the tensor-to-scalar ratio, and,
together with $\eps{2}$, fixes the spectral index (first order). At
second order, $\eps{3}$ makes corrections to the former quantities and
enters into the running of the spectral index, and so on and so forth.

Historically, the second-order corrections for the scalar spectral
index were first derived in Ref.~\cite{Gong:2001he} but the fully
expanded scalar power spectrum at second order (N2LO) was
explicitly derived in Ref.~\cite{Schwarz:2001vv}. The tensor-mode
power spectrum at N2LO was derived soon after in
Ref.~\cite{Leach:2002ar}. For inflation with a non-minimal kinetic term,
the N2LO slow-roll scalar and tensor spectra have been calculated in
Refs.~\cite{Martin:2013uma, Jimenez:2013xwa}. Third-order corrections
to the scalar amplitude, for a minimal kinetic term, were first
derived in Ref.~\cite{Choe:2004zg}, but not fully expanded around a
pivot wavenumber.

The accuracy of the slow-roll expanded power spectra can be assessed
by comparing their predictions to a full numerical integration of the
linear perturbations~\cite{Salopek:1988qh,
  Adams:2001vc, Ringeval:2005yn, Makarov:2005uh, Martin:2006rs,
  Ringeval:2007am, Martin:2010kz, Mortonson:2010er}. Moreover, as
discussed in Refs.~\cite{Leach:2002ar,Martin:2006rs}, performing
robust data analysis of slow-roll inflation requires marginalizing
over the unconstrained higher order slow-roll parameters. These tests
have been done in Ref.~\cite{Ringeval:2013lea} for the Planck data,
and, performing inflationary Bayesian parameter estimation based on
N2LO power spectra ends up being indistinguishable from an exact
numerical treatment. Therefore, up to now, N2LO corrections are
sufficient.

But the situation is bound to change with the next generation of CMB
observations and the incoming large-scale structures surveys. Combined
together, data from the CMB-S4 telescopes and from the LiteBird
satellite are not only going to increase the sensitivity to $B$-modes,
and thus to primordial gravitational waves ($\eps{1}$), but also to
probe much higher multipoles~\cite{CMB-S4:2016ple,
  SimonsObservatory:2018koc, Mallaby-Kay:2021tuk,
  LiteBIRD:2022cnt}. The same remark concerns the Euclid satellite,
and other ground surveys, that will be measuring the small scales in
the matter power spectrum~\cite{LSSTScience:2009jmu, Lacasa:2019flz,
  Euclid:2021qvm}. All these cosmological measurements are going to
increase the lever arm between the large and small angular scales, and,
assuming that foregrounds will be well enough understood to be
removed, or fitted, one should expect the data to be sensitive to any
possible running of the primordial spectral index; namely, to the third
slow-roll parameter
$\eps{3}$~\cite{Adshead:2010mc,Martin:2014rqa,Sprenger:2018tdb}. Let
us also notice that if inflation proceeds at very low-energy scales,
primordial gravitational waves would be undetectable and the first
parameter $\eps{1}$ would end up being very
small~\cite{Ringeval:2010hf, BeltranJimenez:2009oao,
  Glavan:2017jye}. In that situation, at N2LO, the scalar power
spectrum is driven by only $\eps{2}$ and $\eps{3}$, with $\eps{2}
\simeq 1 - \nS \simeq 0.04$, and all new data will necessarily
contribute to an information gain on $\eps{3}$~\cite{Martin:2016oyk,
  Easther:2021eje}.

For all these reasons, in this paper, we push the slow-roll
calculations one step further, and derive at
next-to-next-to-next-to-leading order (N3LO) both the scalar and
tensor power spectrum, with minimal and non-minimal kinetic terms. Our
results are derived using the Green's function
method~\cite{Gong:2001he} and are fully expanded around a pivot
wavenumber. As such, $\eps{4}$ explicitly corrects the lower
perturbative terms and enters into the expression of the running of the
running of the spectral index. The expressions that we derive in this
work will allow us to robustly search for a non-vanishing $\eps{3}$
in the future cosmological data.

Finally, let us clarify that the slow-roll calculations in general,
and the ones presented below, are semi-classical. We are dealing with
quantum fluctuations arising around a classical quasi-de Sitter
spacetime. As such, one cannot go forever including higher-order
semi-classical corrections, at some point the quantum backreaction is
expected to dominate. The latter can actually be estimated using the
stochastic inflation formalism~\cite{Starobinsky:1986fx,
  Starobinsky:1994bd}. The non-perturbative quantum corrections to the
scalar amplitude are given by $\delta_\uqu \calPz/\calPz = \calPz
(\eps{1} + \eps{2}) \simeq 8 \times
10^{-11}$~\cite{Vennin:2020kng}. The semi-classical slow-roll
corrections at order $p$ contain a term of order $\eps{2}^p$, which,
at N3LO, is $\eps{2}^3 \simeq 6 \times 10^{-5}$, i.e., six orders of
magnitude larger than the expected quantum
backreaction\footnote{Non-perturbative quantum corrections to the
spectral index can also be explicitly derived within the stochastic
inflation formalism. One finds that they are accordingly reduced by
one power in the slow-roll parameters. Ignoring terms containing
$\eps{1}$ and $\eps{3}$, one has $\delta_\uqu \nS/\nS \simeq -3 \calPz
\eps{2}^2$.}. This simple estimate suggests that semi-classical
slow-roll corrections are expected to be relevant up to order
six. Notice, however, that the aforementioned reasoning assumes that
quantum diffusion never dominates the field dynamics in the last $60$,
or so, e-folds of inflation. As such, it may not apply if a phase of
quantum diffusion takes place just before the end of inflation, as one
that could produce primordial black holes~\cite{Pattison:2017mbe,
  Carr:2019hud, Auclair:2020csm, Tada:2021zzj}. In that situation, one
can show that the relation between observable wavenumbers and field
values at Hubble crossing become probabilistic, and the observed power
spectra may even be severely modified at first
order~\cite{Ando:2020fjm}.

The paper is organized as follows. In \cref{sec:tensor} we briefly
recap the Green's function method and apply it to the equation of
motion of primordial gravitational waves to derive the slow-roll
tensor power spectrum at N3LO. At this order, a new set of definite
integrals need to be calculated and some details on their derivation
are given in \cref{sec:genfunc}. In \cref{sec:scalar}, we derive the
scalar power spectra, for inflation with minimal and non-minimal
kinetic terms, using the mapping method introduced in
Ref.~\cite{Jimenez:2013xwa}. The scalar results have been
cross-checked using a direct calculation which is summarized in the
\cref{sec:details}. For varying speed of sound models, we also
explicitly recast the tensor spectrum at the same pivot as the scalar
spectrum where it depends on the speed of sound
$\cs$~\cite{Martin:2013uma}.  Finally, using the independence of the
spectra with respect to the pivot scale~\cite{Schwarz:2004tz}, we
give, in \cref{sec:pln4lo}, all the power-law indices expanded at one
more order, namely N4LO. We conclude in \cref{sec:conclusion}.

\section{Primordial gravitational waves}
\label{sec:tensor}

Within the theory of cosmological perturbations, the tensor linear
perturbations around a FLRW spatially flat metric verify, in Fourier
space, the equation of motion~\cite{Mukhanov:1990me}
\begin{equation}
\mu'' + \left(k^2 - \dfrac{a''}{a}\right) \mu = 0,
\label{eq:muevol}
\end{equation}
where a prime denotes differentiation with respect to the conformal
time $\eta$. In this equation, $k$ stands for the comoving wavenumber
and we assume a mostly positive metric signature. The mode function
$\mu(\eta,k)$ is related to the traceless and divergenceless tensor
perturbations $h_{ij}(\eta,k)$ by $\mu(\eta,k) \equiv
h_\lambda(\eta,k) a(\eta)$ where $h_\lambda$ is one of the two
polarization states of $h_{ij}$ ($\lambda=+,\times$). In the
following, the explicit dependence in $\lambda$ is dropped as, within
General Relativity, the two states are identically created and
propagated~\cite{Alexander:2004wk}. They are accounted for, as two
degrees of freedom, when estimating the tensor power
spectrum. Equation~\eqref{eq:muevol} describes a parametric oscillator
evolving in a time-dependent effective potential
\begin{equation}
\Ut(\eta) \equiv \dfrac{a''}{a} = \calH^2 \left( 2- \eps{1} \right),
\label{eq:Utdef}
\end{equation}
where $\calH(\eta) = a(\eta)H(\eta)$ is the conformal Hubble parameter
and use has been made of \cref{eq:hubbleflow}.

\subsection{Hubble flow expansion}

\label{sec:hubbleflow}

The Hubble flow functions introduced in \cref{eq:hubbleflow} can
be used to provide a perturbative expression for the potential
$\Ut$. In the following, we choose the conformal time $\eta$ to be
negative during inflation. From the definition of $\eta$, one gets
\begin{equation}
\begin{aligned}
  -\eta = \int_\eta^{0{^-}} \dfrac{\dd{t}}{a(t)} & =
\int_{a(\eta)}^{+\infty} \dfrac{\dd{a}}{a^2 H(a)} \\ & = \dfrac{1}{a H} +
\int_{a}^{+\infty} \dfrac{1}{a} \dv{\left(H^{-1}\right)}{a} \dd{a},
\end{aligned}
\label{eq:etadef}
\end{equation}
where we have made a change of variable from $t$ to $a(t)$ and
performed an integration by part. The integrand is now proportional to
the derivative of the Hubble radius, i.e., a small quantity during
slow-roll inflation. From \cref{eq:hubbleflow} one has
\begin{equation}
\dv{\left(H^{-1} \right)}{a} = \dfrac{\eps{1}}{a H}\,,
\end{equation}
and \cref{eq:etadef} can be once more integrated by parts as
\begin{equation}
-\eta = \dfrac{1+\eps{1}}{aH} + \int_a^{+\infty} \dfrac{1}{a}
\dv{a}\left(\dfrac{\eps{1}}{H}\right) \dd{a}.
\label{eq:etaparts}
\end{equation}
From \cref{eq:hubbleflow}, the Hubble flow functions verify
\begin{equation}
\dv{\eps{i}}{a} = \dfrac{\eps{i} \eps{i+1}}{a}\,,
\end{equation}
and each successive integration by parts of \cref{eq:etaparts}
ends up being a perturbative expansion in the Hubble flow
functions. For our purpose, this needs to be pushed up to third order
and one gets
\begin{equation}
\begin{aligned}
  -\eta & = \dfrac{1}{\calH}\left(1 + \eps{1} + \eps{1}^2 +
  \eps{1}\eps{2} + \eps{1}^3 + 3 \eps{1}^2 \eps{2} + \eps{1}\eps{2}^2
  + \eps{1} \eps{2} \eps{3} \right) \\ & + \int_a^{+\infty}
  \dfrac{1}{a} \dv{a} \left[\dfrac{1}{H} \left(\eps{1}^3 + 3 \eps{1}^2
    \eps{2} + \eps{1} \eps{2}^2 + \eps{1}\eps{2}\eps{3} \right)\right]
  \dd{a},
\end{aligned}
\label{eq:eta3rd}
\end{equation}
where the integral in the second line is of order $\order{\eps{}^4}$
and can be omitted.

From \cref{eq:Utdef,eq:eta3rd}, we obtain the
perturbative expression for the effective potential $\Ut$
\begin{equation}
\begin{aligned}
  \eta^2 \Ut(\eta) & = 2 + 3 \eps{1} + 4 \eps{1}^2 + 4 \eps{1} \eps{2} + 5
\eps{1}^3 + 14 \eps{1}^2 \eps{2} + 4 \eps{1} \eps{2}^2 \\ & + 4 \eps{1}
\eps{2} \eps{3} + \order{\eps{}^4}.
\end{aligned}
\label{eq:Ut3rd}
\end{equation}
Notice that all $\eps{i}(\eta)$ are function of $\eta$, and, from now
on, we assume that they are all of the same order $\order{\eps{}}$
and no additional assumptions will be made. Nevertheless, in order to
solve \cref{eq:muevol}, one needs the explicit conformal time
dependence of $\Ut$; namely, how the flow functions $\eps{i}$ depend on
$\eta$. In fact, this is already encoded in their
definition. Remarking that $N=\ln(a)$ implies
\begin{equation}
\dv{N}{\ln \abs{\eta}} = \eta \calH,
\label{eq:Nprime}
\end{equation}
one has, from \cref{eq:hubbleflow,eq:eta3rd},
\begin{equation}
\begin{aligned}
  \dv{\eps{i}}{\ln \abs{\eta}} &= \eta \calH  \eps{i}\eps{i+1}
 = - \eps{i} \eps{i+1} \left(1 + \eps{1} + \eps{1}^2 + \eps{1} \eps{2} +
  \eps{1}^3 \right. \\ & \left. + 3 \eps{1}^2 \eps{2} + \eps{1} \eps{2}^2 + \eps{1}
  \eps{2} \eps{3} \right) + \order{\eps{}^6}.
\end{aligned}
\end{equation}
At any order, the derivatives of all the $\eps{i}$ are
polynomials of the $\eps{i}$. One can now pick a peculiar time, say
$\etak$, and Taylor expand the Hubble flow functions at $\etak$. As we will
see later on, the N3LO scalar power spectrum requires determining
$\eps{1}$ up to the fourth order. At this order, the Taylor expansion
of the Hubble flow functions reads
\begin{equation}
\begin{aligned}
 \eps{i}(\eta) & = \epsk{i} - \epsk{i}\epsk{i+1} \left( 1+\epsk{1} +
 \epsk{1}^2 + \epsk{1}\epsk{2} \right)
 \ln\left(\dfrac{\eta}{\etak}\right) \\ & +
 \dfrac{\epsk{i}\epsk{i+1}}{2} \left(\epsk{i+1} + \epsk{i+2} +
 \epsk{1} \epsk{2} + 2 \epsk{1}\epsk{i+1} \right. \\ & \left. + 2
 \epsk{1}\epsk{i+2}\right) \ln^2\left(\dfrac{\eta}{\etak}\right) -
 \dfrac{\epsk{i}\epsk{i+1}}{6} \left(\epsk{i+1}^2 + \epsk{i+2}^2
 \right. \\ & \left. + 3 \epsk{i+1} \epsk{i+2} + \epsk{i+2} \epsk{i+3}
 \right) \ln^3\left(\dfrac{\eta}{\etak}\right) + \order{\eps{}^5},
\end{aligned}
\label{eq:epssr4}
\end{equation}
where we have used the shortcut notation $\epsk{i} \equiv
\eps{i}(\etak)$. All these quantities are now truly
slow-roll ``parameters''. As a result, the effective potential $\Ut$ in
\cref{eq:Ut3rd} inherits an explicit time dependence involving up to the
third power of $\ln(\eta/\etak)$. This allows us to solve
\cref{eq:muevol} by the Green's function method~\cite{Gong:2001he}.

\subsection{Green's function}

Using the dimensionless positive variable $x\equiv -k \eta$,
\cref{eq:muevol} can be recast into
\begin{equation}
\dv[2]{\mu}{x} + \left[1 - \dfrac{\Ut(x)}{k^2}\right] \mu
= 0,
\end{equation}
which, from \cref{eq:Ut3rd,eq:epssr4}, can be split
as~\cite{Gong:2001he}
\begin{equation}
\dv[2]{\mu}{x} + \left(1 - \dfrac{2}{x^2} \right) \mu =
\dfrac{g[\ln(x)]}{x^2} \mu.
\label{eq:musplit}
\end{equation}
By choosing a convenient conformal time
\begin{equation}
\etak = -\dfrac{1}{k}\,,
\label{eq:etak}
\end{equation}
the function $g$ reduces to a polynomial in $\ln(x)$. At third order in
the slow-roll parameters, it reads
\begin{equation}
g[\ln(x)] = \gk{1} + \gk{2} \ln(x) + \gk{3} \ln^2(x) + \order{\eps{}^4},
\label{eq:gfunc}
\end{equation}
with
\begin{equation}
  \begin{aligned}
    \gk{1} & = 3 \epsk{1} + 4 \epsk{1}^2 + 4 \epsk{1} \epsk{2} + 5
    \epsk{1}^3 + 14 \epsk{1}^2 \epsk{2} + 4 \epsk{1} \epsk{2}^2  \\ & + 4
    \epsk{1} \epsk{2} \epsk{3}+
    \order{\eps{}^4}, \\
    \gk{2} & = -\left( 3\epsk{1} \epsk{2} + 11 \epsk{1}^2\epsk{2} + 4
    \epsk{1}\epsk{2}^2 + 4 \epsk{1} \epsk{2} \epsk{3} \right) \\ & +
    \order{\eps{}^4}, \\
    \gk{3} & = \dfrac{3}{2} \left(\epsk{1}\epsk{2}^2 + \epsk{1}
    \epsk{2} \epsk{3} \right)+
    \order{\eps{}^4}.
  \end{aligned}
\label{eq:gi}
\end{equation}
The left-hand side of \cref{eq:musplit} is the equation of
gravitational waves (GW) propagating in a pure matter era, namely a
Riccati-Bessel equation~\cite{Abramovitz:1970aa, daCunha:2021wyy}. As
such, \cref{eq:musplit} can be formally solved using the advanced\footnote{We
have defined $x=-k \eta$ and this corresponds to retarded solution in
$\eta$.} Green's function sourced by a Dirac distribution $\delta(x-y)$. The
boundary conditions are chosen to match positive energy waves at small
scales, i.e., varying as $e^{-ik\eta}$ for $k \eta \to -\infty$. From
the Wronskian method, one gets
\begin{equation}
\G{y}(x) = \dfrac{i}{2} \left[u(x) \conj{u}(y) - u(y)
  \conj{u}(x)\right] \heaviside{y-x},
\end{equation}
where $u(x)$ is the Riccati-Hankel function of order one:
\begin{equation}
u(x) \equiv \left(1+ \dfrac{i}{x}\right)e^{i x},
\label{eq:ufunc}
\end{equation}
and $\conj{u}$ denotes its complex conjugate. The exact solution of
\cref{eq:musplit} reads
\begin{equation}
\begin{aligned}
  \mu(x) & = \dfrac{i}{2} \int_x^\infty \dfrac{g[\ln(y)]}{y^2} \mu(y)
  \left[\conj{u}(y) u(x) -\conj{u}(x) u(y)\right]\dd{y} \\ & +
  \mu_0(x),
\end{aligned}
\label{eq:greensol}
\end{equation}
with $\mu_0(x)$ a solution of the homogeneous equation. From
\cref{eq:gi}, we see that the leading-order term in the function
$g[\ln(x)]$ is of order $\order{\eps{}}$ and \cref{eq:greensol} is
readily a perturbative expansion in the Hubble flow
functions. Imposing the Bunch-Davies vacuum state for $k \eta \to -
\infty$, the zero-order term is completely
fixed~\cite{Mukhanov:1990me}. In Planck units, one gets
\begin{equation}
\mu_0(x) = \dfrac{u(x)}{\sqrt{k}}\,.
\label{eq:mu0}
\end{equation}
Defining the rescaled mode function
\begin{equation}
\muh{}(x) \equiv \sqrt{k} \mu(x),
\end{equation}
we can expand it as
\begin{equation}
\muh{}(x) = \muh{0}(x) + \muh{1}(x) + \muh{2}(x) + \muh{3}(x) + \order{\eps{}^4},
\label{eq:musr3}
\end{equation}
with $\muh{0}(x) = u(x)$. From \cref{eq:greensol,eq:musr3}, one finally gets the recursive solutions
\begin{align}
\label{eq:mu1}
  \muh{1}(x) & = \gk{1} \int_x^{\infty} \dfrac{\G{y}(x)}{y^2} u(y)
  \dd{y}, \\
\label{eq:mu2}
  \muh{2}(x) & =\gk{2} \int_x^\infty \dfrac{\G{y}(x)\ln(y)}{y^2} u(y)
 \nonumber \\ & + \gk{1} \int_x^\infty \dfrac{\G{y}(x)}{y^2} \muh{1}(y) \dd{y},
\end{align}
and
\begin{equation}
\begin{aligned}
\muh{3}(x) & = \gk{3} \int_x^\infty \dfrac{\G{y}(x) \ln^2(y)}{y^2} u(y)
\dd{y} \\ & + \gk{2} \int_x^\infty \dfrac{\G{y}(x) \ln(y)}{y^2}\muh{1}(y)
\dd{y} \\ & + \gk{1} \int_x^\infty \dfrac{\G{y}(x)}{y^2} \muh{2}(y) \dd{y},
\end{aligned}
\label{eq:mu3}
\end{equation}
which consistently ensure that $\muh{p} = \order{\eps{}^p}$.

\subsection{Solution at N3LO}

Pushing the slow-roll expansion at N3LO not only complexifies the
functional form of the $\gk{i}$ parameters but also introduces new
definite multidimensional integrals. For instance, the last term of
\cref{eq:mu3} is an integral over $\muh{2}(y)$, which is itself
defined by integrals over $\muh{1}(y)$ and $u(y)$. Using the explicit
form of $u(y)$ given in \cref{eq:ufunc} and recursing on
integration by parts, one can show that \cref{eq:mu1,eq:mu2,eq:mu3} are completely determined by a few integrals that we
now enumerate.

One has first a family of one-dimensional integrals
\begin{equation}
\F{n}(x) \equiv \int_x^\infty \dfrac{e^{+2 i y}}{y} \ln^n(y) \dd{y},
\label{eq:Fn}
\end{equation}
for $n=0$, $n=1$ and $n=2$. These then enter into the definition of three
other two-dimensional integrals
\begin{equation}
\begin{aligned}
  \F{00}(x) &= \int_x^\infty \dfrac{e^{-2iy}}{y} \F{0}(y) \dd{y},\\
  \F{01}(x) & = \int_x^\infty \dfrac{e^{-2iy}}{y} \F{1}(y) \dd{y},\\
  \F{10}(x) & = \int_x^\infty  \dfrac{e^{-2iy}}{y} \ln(y) \F{0}(y)\dd{y}.
\end{aligned}
\label{eq:Fnp}
\end{equation}
Finally, the third-order terms involve the three-dimensional integral
\begin{equation}
\F{000}(x) = \int_x^\infty \dfrac{e^{+2iy}}{y}\F{00}(y)\dd{y}.
\label{eq:F000}
\end{equation}

In terms of these integrals, the explicit solution for $\muh{1}$ reads
\begin{equation}
\muh{1}(x) = -\dfrac{{\left[{\left(x - i\right)} \F{0}(x) -
    2 i \, e^{2 i x}\right]} \gk{1} e^{-i x }}{3 x}\,,
\end{equation}
while the second order term simplifies to
\begin{widetext}
\begin{equation}
  \begin{aligned}
    \muh{2}(x) & = \frac{\gk{1}^2}{27} \left[3  \F{00}(x) e^{i x} + \F{0}(x) e^{-i  x} +
      \frac{3 i \F{00}(x) e^{i x}}{x} +
      \frac{5 i \F{0}(x) e^{-i x}}{x} - \frac{2 i  e^{i x}}{x}\right]  \\
      & -
    \frac{ \gk{2}}{9} \left[\vphantom{\frac{e^{i x}}{x}} 7 \F{0}(x) e^{-i x} + 3
      \F{1}(x) e^{-i x}
      - \frac{7 i \F{0}(x) e^{-i x}}{x}- \frac{3 i \F{1}(x) e^{-i x}}{x}
      - \frac{6 i e^{ix} \ln(x)}{x} -
      \frac{8 i e^{i x}}{x}\right].
  \end{aligned}
\end{equation}
Most of the action takes place for the third order correction which,
after some algebra, reads
\begin{equation}
\begin{aligned}
 \muh{3}(x) &= -\frac{\gk{1}^3}{243} \left[6 \F{00}(x) e^{i
     x} + 2 \F{0}(x) e^{-i x} + 9
   \F{000}(x) e^{-i x} - \frac{12 i
     \F{00}(x) e^{i x}}{x} + \frac{10 i
     \F{0}(x) e^{-i x}}{x} - \frac{9 i
     \F{000}(x) e^{-i x}}{x} \right. \\ &  \left. - \frac{4 i
     e^{i x}}{x}\right]
 + \frac{\gk{1} \gk{2}}{81}
     \left[21 \F{00}(x) e^{i x} + 9
       \F{01}(x) e^{i x} + 26
       \F{0}(x) e^{-i x} + 3
       \F{1}(x) e^{-i x} + \frac{21 i
         \F{00}(x) e^{i x}}{x} + \frac{9 i
         \F{01}(x) e^{i x}}{x} \right. \\ & \left. + \frac{16 i
         \F{0}(x) e^{-i x}}{x} + \frac{15 i
         \F{1}(x) e^{-i x}}{x} - \frac{6 i
         e^{i x} \ln(x)}{x} - \frac{28 i
         e^{i x}}{x}\right]
     + \frac{\gk{1} \gk{2}}{81}
     \left[\vphantom{\frac{e^{i x}}{x}} 21 \F{00}(x) e^{i x} + 9
       \F{10}(x) e^{i x} \right. \\ & \left. + 26
       \F{0}(x) e^{-i x} + 3
       \F{1}(x) e^{-i x} + \frac{18 i
         \F{0}(x) e^{-i x}
         \ln(x)}{x} + \frac{21 i \F{00}(x)
         e^{i x}}{x} + \frac{9 i \F{10}(x)
         e^{i x}}{x} - \frac{2 i \F{0}(x)
         e^{-i x}}{x} \right. \\ & \left. - \frac{3 i \F{1}(x)
         e^{-i x}}{x} - \frac{6 i e^{i x}
         \ln(x)}{x} - \frac{28 i e^{i x}}{x}\right]
     - \frac{\gk{3}}{27}
     \left[50 \F{0}(x) e^{-i x} + 42
       \F{1}(x) e^{-i x} + 9
       \F{2}(x) e^{-i x} - \frac{18 i
         e^{i x} \ln(x)^2}{x} \right. \\ & \left. - \frac{50 i
         \F{0}(x) e^{-i x}}{x} - \frac{42 i
         \F{1}(x) e^{-i x}}{x} - \frac{9 i
         \F{2}(x) e^{-i x}}{x} - \frac{48 i
         e^{i x} \ln(x)}{x} - \frac{52 i
         e^{i x}}{x}\right].
\end{aligned}
\end{equation}
\end{widetext}
Let us notice that the time dependence in $x$ appears explicitly in
functions such as $\ln(x)$, but also implicitly through the integrals
introduced earlier. Most of these integrals do not have an explicit
expression, but, as we show in \cref{sec:genfunc}, they can be
recursively derived from a generating functional and this will allow us to
uniquely determine their super-Hubble limit. From the theory of the
cosmological perturbations, we know that, after Hubble exit, for $x
\ll 1$, $\mu/a$ should be conserved and can no longer depend on
$x$. As a result, a consistency check of the whole calculation is that
all the terms in $\ln(x)$, which are singular in the limit $x \to 0$,
must cancel in the final result.

\subsection{Power spectrum}

The constancy of $\mu/a$ after Hubble exit allows us to derive the
observable power spectrum of gravitational waves generated during
inflation. Taking into account the polarization degrees of freedom,
one has
\begin{equation}
\calPh(k) = \dfrac{2 k^3}{\pi^2} \lim_{x\to 0} \left|\dfrac{\mu}{a}\right|^2,
\label{eq:calPhdef}
\end{equation}
and we need the explicit expression of $a(\eta)$, at third order in the
Hubble flow functions. This one can be obtained from the definition of
the number of e-folds. From \cref{eq:Nprime}, one obtains
\begin{equation}
\begin{aligned}
  \Delta\Nk & \equiv N - \Nk = -\left(1+\epsk{1} + \epsk{1}^2 +
  \epsk{1} \epsk{2} + \epsk{1}^3 \right. \\ & \left. + 3 \epsk{1}^2
  \epsk{2} + \epsk{1} \epsk{2}^2 + \epsk{1} \epsk{2} \epsk{3} \right)
  \ln\left(\dfrac{\eta}{\etak}\right) \\ & +
  \dfrac{1}{2}\left(\epsk{1} \epsk{2} + 3 \epsk{1}^2 \epsk{2} +
  \epsk{1} \epsk{2}^2 + \epsk{1} \epsk{2} \epsk{3} \right) \ln^2
  \left(\dfrac{\eta}{\etak}\right) \\ & - \dfrac{1}{6}
  \left(\epsk{1}\epsk{2}^2 + \epsk{1}\epsk{2} \epsk{3} \right) \ln^3
  \left(\dfrac{\eta}{\etak}\right) + \order{\eps{}^4},
\end{aligned}
\end{equation}
from which we get
\begin{equation}
  a(\eta) = \ak e^{\Delta\Nk}.
\end{equation}
As the scale factor is not a measurable quantity, we can trade $\ak$
for $\Hk$ by remarking that $\ak = \calHk/\Hk$ and making use of
\cref{eq:eta3rd}. One finally gets
\begin{widetext}
\begin{equation}
  \begin{aligned}
    a(\eta) & =  \dfrac{-1}{\eta \Hk} \bigg[ 1 + \epsk{1} + \epsk{1}^2 + \epsk{1} \epsk{2}
    + \epsk{1}^3 + 3\epsk{1}^2 \epsk{2}  + \epsk{1} \epsk{2}^2 +
    \epsk{1} \epsk{2} \epsk{3} - \left(\epsk{1} + 2 \epsk{1}^2 + \epsk{1}\epsk{2}  + 3 \epsk{1}^3 + 5 \epsk{1}^2 \epsk{2}
    \right. \\ &+ \left. \epsk{1}\epsk{2}^2 + \epsk{1}\epsk{2} \epsk{3} \right)
    \ln\left(\dfrac{\eta}{\etak}\right) + \dfrac{1}{2}
    \left(\epsk{1}^2 + \epsk{1} \epsk{2} + 3 \epsk{1}^3 + 6 \epsk{1}^2
    \epsk{2} +  \epsk{1} \epsk{2}^2  +  \epsk{1} \epsk{2} \epsk{3}
    \right) \ln^2\left(\dfrac{\eta}{\etak}\right) \\ & - \dfrac{1}{6}
    \left(\epsk{1}^3 + 3 \epsk{1}^2 \epsk{2}  + \epsk{1} \epsk{2}^2 +
    \epsk{1} \epsk{2} \epsk{3} \right)
    \ln^3\left(\dfrac{\eta}{\etak}\right) + \order{\eps{}^4} \bigg] .
  \end{aligned}
\label{eq:ahat}
\end{equation}
This expression suggests defining the rescaled quantity
\begin{equation}
\ah \equiv \dfrac{a \Hk x}{k}\,,
\end{equation}
which is an explicit function of $\ln(x)$ only. From
\cref{eq:calPhdef}, one has
\begin{equation}
\calPh(k) = \lim_{x\to 0} \dfrac{2 \Hk^2}{\pi^2} \left| \dfrac{x \muh{}}{\ah}\right|^2,
\label{eq:calPhx}
\end{equation}
which can be evaluated using the expression $\ah$ in
\cref{eq:ahat}. Some intermediate steps are given in the
\cref{sec:details} and, after some serious algebra, keeping
only the leading-order terms of all the integrals $\F{n}(x)$ and
$\F{0^n}(x)$, the divergent terms in $\ln(x)$ all cancel. Dropping the
explicit order at which the slow-roll expansion is made, the
super-Hubble limit finally reads
\begin{equation}
\begin{aligned}
  \lim_{x \to 0} \left| \dfrac{x \muh{}}{\ah}\right|^2 & = 1 - 2(1+C)\epsk{1} +
  \frac{1}{2} \left(\pi^2 + 4 C^2 + 4 C - 6\right)
  \epsk{1}^2 + \frac{1}{12} \left(\pi^2 - 12 C^2 - 24 C
  - 24\right) \epsk{1} \epsk{2} \\ &-\frac{1}{3}
  \left[4 C^3 + 3 {\left(\pi^2 - 8\right)} C + 14 \zeta(3) -
  16\right] \epsk{1}^3 + \frac{1}{12} \left[24 C^3 +
      13 \pi^2 + 2 {\left(5 \pi^2 - 36\right)} C + 36 C^2 -
      96\right] \epsk{1}^2 \epsk{2} \\ & -
    \frac{1}{12} \left[4 C^3 - \pi^2 - {\left(\pi^2 -
        24\right)} C + 12 C^2 + 8 \zeta(3) + 8\right]
         \left( \epsk{1} \epsk{2}^2 + \epsk{1}
         \epsk{2} \epsk{3} \right).
\end{aligned}
\label{eq:xmuoah}
\end{equation}
\end{widetext}
In this expression, $\zeta(3)$ is a number given by the Riemann zeta
function $\zeta(z)$ evaluated at $z=3$. Similarly, $C$ stands for
another number
\begin{equation}
C \equiv \gammaE + \ln(2) - 2,
\end{equation}
$\gammaE$ being the Euler constant. The dependence of
\cref{eq:xmuoah} in the wavenumber $k$ is implicit and hidden by
our choice of the peculiar time $\etak = -1/k$ at which all the Hubble
flow functions are evaluated. In order to make contact with the
cosmological observations, one needs to recast this expression at a
fixed wavenumber, say $\kast$, chosen by the observer. For CMB
measurements, the typical wavenumber would be $\kast =
0.05\,\Mpc^{-1}$~\cite{Planck:2018nkj} and this implies that the
Hubble flow functions should be expanded around another time uniquely
fixed by $\kast$. Let us choose a time, say $\etaast$, close to the Hubble
crossing and verifying
\begin{equation}
\kast \etaast  = -1.
\label{eq:etaast}
\end{equation}
Using the Hubble flow expansions of \cref{eq:epssr4}, we can expand
all parameters $\epsk{i}$ in terms of $\epsast{i}=\eps{i}(\etaast)$
and powers of
\begin{equation}
\ln\left(\dfrac{\etak}{\etaast}\right) = -\ln\left(\dfrac{k}{\kast}\right).
\end{equation}
Let us stress that the normalization $\Hk^2$ appearing in
\cref{eq:calPhx} has to be expanded around the new time
$\etaast$. This can be straightforwardly done from
\cref{eq:eta3rd} by remarking that
\begin{equation}
\dv{\ln H}{\ln \abs{\eta}} = -\eps{1} \eta \calH.
\end{equation}
After some algebra, one finally obtains one of the main results of
this work, namely, the tensor mode power spectrum, at N3LO, fully
expanded around the pivot wavenumber $\kast$. It reads
\begin{widetext}
\begin{equation}
\begin{aligned}
    \calPh(k) & = \dfrac{2 \Hast^2}{\pi^2} \bigg\{
        1 - 2 \qty(C + 1) \epsast{1}
        + \frac{1}{2} \qty(\pi^2 + 4 C^2 + 4 C - 6) \epsast{1}^2
        + \frac{1}{12} (\pi^2 - 12 C^2 - 24 C - 24) \epsast{1} \epsast{2} \\
        & -\frac{1}{3} \qty[4 C^3 + 3 (\pi^2 - 8) C + 14 \zeta(3) - 16] \epsast{1}^3
        + \frac{1}{12} \qty[24 C^3 + 13 \pi^2 + 2 (5 \pi^2 - 36) C + 36 C^2 - 96] \epsast{1}^2 \epsast{2} \\
        & - \frac{1}{12} \qty[4 C^3 - \pi^2 - (\pi^2 - 24) C + 12 C^2 + 8 \zeta(3) + 8] (\epsast{1} \epsast{2}^2 + \epsast{1} \epsast{2} \epsast{3}) \\
        & + \bigg[
            - 2 \epsast{1}
            + 2 (2 C + 1) \epsast{1}^2
            - 2 (C + 1) \epsast{1} \epsast{2}
            - (\pi^2 + 4 C^2 - 8) \epsast{1}^3
            + \frac{1}{6} (5 \pi^2 + 36 C^2 + 36 C - 36) \epsast{1}^2 \epsast{2} \\
            & + \frac{1}{12} (\pi^2 - 12 C^2 - 24 C - 24) (\epsast{1} \epsast{2}^2 + \epsast{1} \epsast{2} \epsast{3})
        \bigg] \ln \qty(\frac{k}{\kast}) \\
        & + \bigg[
            2 \epsast{1}^2
            - \epsast{1}\epsast{2}
            - 4 C \epsast{1}^3
            + 3 (2 C + 1) \epsast{1}^2\epsast{2}
            - (C + 1) (\epsast{1} \epsast{2}^2 + \epsast{1} \epsast{2} \epsast{3})
        \bigg] \ln^2 \qty(\frac{k}{\kast}) \\
        & + \frac{1}{3} \bigg(
            - 4 \epsast{1}^3
            + 6 \epsast{1}^2 \epsast{2}
            - \epsast{1} \epsast{2}^2
            - \epsast{1} \epsast{2} \epsast{3}
        \bigg) \ln^3 \qty(\frac{k}{\kast})
    \bigg\}.
  \end{aligned}
\label{eq:calPhsr3ast}
\end{equation}
\end{widetext}
This formula corrects the one derived independently in an unpublished
work~\cite{Beuseling:2019} (master thesis).

\section{Curvature perturbations}
\label{sec:scalar}

The dominant anisotropies in the universe are sourced by scalar
perturbations and, in this section, we derive the primordial power
spectrum of the comoving curvature perturbation $\zeta(\eta,k)$
\begin{equation}
\calPz(k) \equiv \dfrac{k^3}{2\pi^2} \left|\zeta\right|^2.
\end{equation}
As a warm-up, we start with an inflationary era driven by a scalar
field having a minimal kinetic term before turning to the more exotic
situation of K-inflation.

\subsection{Minimal kinetic term}

\subsubsection{Equation of motion}

The evolution of the curvature perturbations during inflation can be
recast as a parametric oscillator by introducing the Mukhanov-Sasaki
variable
\begin{equation}
v(\eta,k) \equiv a(\eta) \sqrt{2\eps{1}(\eta)} \zeta(\eta,k).
\end{equation}
In Fourier space, it satisfies~\cite{Mukhanov:1990me}
\begin{equation}
\dv[2]{v}{\eta} + \left(k^2 - \dfrac{1}{z}\dv[2]{z}{\eta} \right)v = 0,
\label{eq:vevol}
\end{equation}
where we have defined
\begin{equation}
z(\eta) \equiv a(\eta)\sqrt{\eps{1}(\eta)}\,.
\end{equation}
Since this equation is formally identical to the one of the tensor modes,
see \cref{eq:muevol}, one could repeat the calculations done in
\cref{sec:tensor}. We provide a summary of these calculations in
\cref{sec:sdetails}, while here, we give the result using
the mapping method discussed in Ref.~\cite{Jimenez:2013xwa}.

\subsubsection{Mapping method}
\label{sec:mapping}

This method consists in introducing a generalized conformal Hubble
parameter $\calHg$ and a generalized e-fold number $\Ng$ such that
\begin{equation}
\calHg \equiv \dfrac{z'}{z} = \dv{\Ng}{\eta}\,.
\label{eq:Hgdef}
\end{equation}
In other word, the function $z(\eta)$ is viewed as a generalized scale
factor. Defining $\Hg \equiv \calHg/z$, one can define a hierarchy of
generalized flow functions
\begin{equation}
\alp{i+1}(\Ng) \equiv \dv{\ln\left|\alp{i}\right|}{\Ng}\,, \qquad
\alp{1}(\Ng) \equiv -\dv{\ln\Hg}{\Ng}\,.
\label{eq:alpdef}
\end{equation}
Let us notice that, although not explicitly needed, the generalized
cosmic time is then defined as $\dd{\tg} = z \dd{\eta}$. Because
\cref{eq:muevol,eq:vevol} are formally identical and have the same
initial conditions (a Bunch-Davies vacuum), up to an overall
normalization accounting for the different number of polarization
degrees of freedom, the scalar power spectrum must be given by
\cref{eq:calPhsr3ast} up to the replacement rule
\begin{equation}
\epsast{i} \to \alpast{i}, \qquad \Hast \to \Hgast.
\end{equation}
The only work to do is expressing the generalized parameters in terms
of the standard Hubble flow functions. From \cref{eq:Hgdef,eq:alpdef}, one has the exact relations
\begin{equation}
\Hg = \dfrac{H}{\sqrt{\eps{1}}} \left(1+ \dfrac{\eps{2}}{2}\right),
\qquad \Ng = N + \dfrac{1}{2} \ln \eps{1},
\label{eq:Hg}
\end{equation}
from which we get~\cite{Jimenez:2013xwa}
\begin{equation}
  \begin{aligned}
    \alp{1} & = \dfrac{4 \eps{1} + \eps{2} \left( 2 + 2\eps{1} +
      \eps{2}-2\eps{3}\right)}{\left(\eps{2}+2\right)^2}\,,\\ \alp{i+1}
    & = \left(1+\dfrac{\eps{2}}{2}\right)
    \dv{\ln\left|\alp{i}\right|}{N}\,.
  \end{aligned}
\label{eq:alp}
\end{equation}
Using \cref{eq:hubbleflow} in this last equation allows us to
determine, by recurrence, all the needed function $\alp{1}$, $\alp{2}$
and $\alp{3}$ in terms of $\eps{i}$ (see Ref.~\cite{Jimenez:2013xwa}).

\subsubsection{Power spectrum}

Plugging \cref{eq:Hg,eq:alp} in the tensor power
spectrum of \cref{eq:calPhsr3ast}, dropping all terms of order equal
or higher than $\order{\eps{}^4}$, and dividing by $16$ to account for
the scalar versus tensor degrees of freedom, we finally obtain
\begin{widetext}
\begin{equation}
\begin{aligned}
\calPz(k) &= \dfrac{\Hast^2}{8 \pi^2 \epsast{1}}\bigg\{
    1 - 2 (C + 1) \epsast{1} - C \epsast{2}
    + \qty(\frac{\pi^2}{2} + 2 C^2 + 2 C - 3) \epsast{1}^2
    + \qty(\frac{7 \pi^2}{12} + C^2 - C - 6) \epsast{1} \epsast{2} \\
    & + \frac{1}{8} \qty(\pi^2 + 4 C^2 - 8) \epsast{2}^2
    + \frac{1}{24} (\pi^2 - 12 C^2) \epsast{2} \epsast{3}
    - \frac{1}{24} \qty[4 C^3 + 3 \qty(\pi^2 - 8) C + 14 \zeta(3) - 16] \qty(8 \epsast{1}^3 + \epsast{2}^3) \\
    & + \frac{1}{12} \qty[13 \pi^2 - 8 (\pi^2 - 9) C + 36 C^2 - 84 \zeta(3)] \epsast{1}^2 \epsast{2}
    - \frac{1}{24} \qty[8 C^3 - 15 \pi^2 + 6 (\pi^2 - 4) C - 12 C^2 + 100 \zeta(3) + 16] \epsast{1} \epsast{2}^2 \\
    & + \frac{1}{24} \qty[\pi^2 C - 4 C^3 - 8 \zeta(3) + 16] \qty(\epsast{2} \epsast{3}^2 + \epsast{2} \epsast{3} \epsast{4})
     + \frac{1}{24} \qty[12 C^3 + \qty(5 \pi^2 - 48) C] \epsast{2}^2
     \epsast{3} \\
     & + \frac{1}{12} \qty[8 C^3 + \pi^2 + 6 \qty(\pi^2 - 12) C - 12 C^2 - 8 \zeta(3) - 8] \epsast{1} \epsast{2} \epsast{3} \\
    & + \bigg[
        - 2 \epsast{1}
        - \epsast{2}
        + 2 (2 C + 1) \epsast{1}^2
        + (2 C - 1) \epsast{1} \epsast{2}
        + C \epsast{2}^2
        - C \epsast{2} \epsast{3}
        - \frac{1}{8} \qty(\pi^2 + 4 C^2 - 8) (8 \epsast{1}^3 + \epsast{2}^3) \\
        & - \frac{2}{3} \qty(\pi^2 - 9 C - 9) \epsast{1}^2 \epsast{2}
        - \frac{1}{4} \qty(\pi^2 + 4 C^2 - 4 C - 4) \epsast{1} \epsast{2}^2
        + \frac{1}{2} \qty(\pi^2 + 4 C^2 - 4 C - 12) \epsast{1} \epsast{2} \epsast{3} \\
        & + \frac{1}{24} \qty(\pi^2 - 12 C^2) \left(\epsast{2} \epsast{3}^2 + \epsast{2} \epsast{3} \epsast{4}\right)
        + \frac{1}{24} \qty(5 \pi^2 + 36 C^2 - 48) \epsast{2}^2 \epsast{3}
    \bigg] \ln\qty(\frac{k}{\kast}) \\
    & + \frac{1}{2} \bigg[
        4\epsast{1}^2
        + 2 \epsast{1} \epsast{2}
        + \epsast{2}^2
        - \epsast{2} \epsast{3}
        + 6 \epsast{1}^2 \epsast{2}
        - (2 C - 1) \qty(\epsast{1} \epsast{2}^2 - 2 \epsast{1} \epsast{2} \epsast{3}) \\
        & - C \left(8\epsast{1}^3
        + \epsast{2}^3
        - 3\epsast{2}^2 \epsast{3}
        + \epsast{2} \epsast{3}^2
        + \epsast{2} \epsast{3} \epsast{4}\right)
    \bigg] \ln^2\qty(\frac{k}{\kast}) \\
    & + \frac{1}{6} \qty(
        - 8 \epsast{1}^3
        - 2 \epsast{1} \epsast{2}^2
        + 4 \epsast{1} \epsast{2} \epsast{3}
        - \epsast{2}^3
        + 3 \epsast{2}^2 \epsast{3}
        - \epsast{2} \epsast{3}^2
        - \epsast{2} \epsast{3} \epsast{4}
    ) \ln^3\qty(\frac{k}{\kast})
    \bigg\}.
\end{aligned}
\label{eq:calPzsr3ast}
\end{equation}
\end{widetext}
As announced in the introduction, let us notice the appearance of
$\epsast{4}$ and an explicit dependence into the third power of
$\ln(k/\kast)$. Both \cref{eq:calPhsr3ast}, and
\cref{eq:calPzsr3ast} determine the tensor and scalar perturbations
generated during single field slow-roll inflation and they are the
main results of this article.

They are, however, not applicable to the more exotic single field
models with varying speed of sound $\cs$. We now turn to this case.

\subsection{K-Inflation}

These models are characterized by a non-minimal kinetic term for the
scalar field $\phi$ while gravity is still given by General
Relativity~\cite{Armendariz-Picon:1999hyi, Garriga:1999vw,
  Shumaylov:2021qje}. Denoting the minimal kinetic term by
\begin{equation}
  X=-\dfrac{1}{2} g^{\mu\nu}\partial_\mu\phi \partial_\nu\phi,
\end{equation}
the Lagrangian density governing the field evolution introduces a new
function $P(X,\phi)$. For a minimal kinetic term, it would be
$P=X-V(\phi)$, $V(\phi)$ being the field's potential, but, here, we
consider it to be a free function. However, not all choices for
$P(X,\phi)$ are acceptable, and, for the equations of motion to remain
hyperbolic, one should have~\cite{Bruneton:2006gf}
\begin{equation}
\pdv{P}{X} > 0, \qquad \pdv{P}{X} + 2 X \pdv[2]{P}{X} > 0.
\end{equation}
Let us notice that these conditions are automatically satisfied when
the non-minimal kinetic term arises as low-dimensional projection of
string theory models, such as in the Dirac--Born--Infeld
inflation~\cite{Alishahiha:2004eh, Ringeval:2009jd}.

\subsubsection{Equation of motion}

At the perturbative level, K-inflation exhibits a speed of sound $\cs$
which does no longer match the speed of light, one has
\begin{equation}
\cs^2 \equiv \dfrac{\displaystyle \pdv{P}{X}}{\displaystyle \pdv{P}{X}
  + 2 X \pdv[2]{P}{X}}\,,
\end{equation}
and this is, in general, a function of the time. As a result, the
sonic radius does no longer match the Hubble radius, and we introduce
a new hierarchy of sound flow functions
\begin{equation}
\del{i+1} \equiv \dv{\ln\abs{\del{i}}}{N}\,,\qquad \del{1} \equiv - \dv{\ln \cs}{N}\,.
\label{eq:soundflow}
\end{equation}
In the following, we assume that all $\del{i}=\order{\eps{}}$ as to
allow for consistent expansions in all the Hubble and sound flow
functions.

The evolution of the comoving curvature perturbation is governed by an
equation formally identical to \cref{eq:vevol},
namely~\cite{Kinney:2007ag}
\begin{equation}
\dv[2]{v}{\tau} + \left(k^2 - \dfrac{1}{z}\dv[2]{z}{\tau}\right)v=0,
\label{eq:nonminimal-eom}
\end{equation}
where we have defined a new time $\tau$
\begin{equation}
    \tau(\eta)  = -\int_\eta^{0^-} \cs(\eta) \dd{\eta},
\label{eq:taudef}
\end{equation}
and the new function
\begin{equation}
  z(\tau) \equiv a(\tau) \sqrt{\dfrac{\eps{1}(\tau)}{\cs(\tau)}}\,.
\label{eq:zkinf}
\end{equation}
The Mukhanov-Sasaki variable remains formally unchanged and reads
\begin{equation}
  v(\tau,k) \equiv
  \sqrt{2} \zeta(\tau,k) z(\tau).
\end{equation}

\subsubsection{Mapping from the tensor modes}
\label{sec:mapping-tensor}

We can again use the mapping method summarized in
\cref{sec:mapping}, constructed on the function $z(\tau)$ defined
in \cref{eq:zkinf}. The generalized conformal Hubble parameter
and e-fold number now read
\begin{equation}
\Hg = \dfrac{H}{\sqrt{\eps{1}}\cs} \left(1+\dfrac{\eps{2}+\del{1}}{2}
\right), \qquad \Ng = N + \dfrac{\ln\eps{1} - \ln \cs}{2}\,,
\end{equation}
from which we get the exact relations
\begin{equation}
\begin{aligned}
  \alp{1} &=
\dfrac{\left(2\eps{1}+\eps{2}-\del{1}\right)\left(\eps{2}+\del{1}+2\right)
  - 2 \left(\del{1}\del{2} + \eps{2} \eps{3}\right)}{\left(\eps{2} + \del{1} + 2\right)^2}\,,\\
\alp{i+1} & = \left(1 + \dfrac{\eps{2}+\del{1}}{2}\right) \dv{\ln\abs{\alp{i}}}{N}\,.
\end{aligned}
\end{equation}

Formally replacing $\Hast \to \Hg$ and $\epsast{i} \to \alp{i}$ into
\cref{eq:calPhsr3ast}, using the above relations and discarding
all terms equal of higher than $\order{\eps{}^4}$, one gets the wanted
expression for $\calPz(k)$, fully expanded around a pivot wavenumber,
say $\ksnd$. This expression can be found in the
\cref{sec:sdetails}, see \cref{eq:calPzsr3snd}, and we do
not repeat it here. The important point is to notice that the above
relations imply that we have also mapped $\eta$ to $\tau$ and the
actual value of $\tausnd$ now verifies
\begin{equation}
\ksnd \tau(\etasnd) = -1.
\label{eq:etasnd}
\end{equation}
It is different from the pivot definition of the tensor modes
($\kast \etaast=-1$). This is problematic because having different
pivots for the scalar and tensor spectra forbids to extract various
observable quantities such as the tensor-to-scalar ratio~\cite{Agarwal:2008ah}.

\subsubsection{Change of pivot}
\label{sec:pivot}
One possibility is to change the scalar mode pivot from $\ksnd
\tau(\etasnd)=-1$ to the one associated with the tensor modes
$\kast\etaast=-1$, and we discuss this possibility later on. However,
because the primordial gravitational waves have not yet been detected,
let us first follow what is usually done and define the pivot using a
characteristic time associated with the scalar perturbations. For
instance, let us define $\etadiam$ when a scalar mode of peculiar
wavenumber $\kdiam$ crossed the sonic radius and became frozen:
\begin{equation}
\kdiam \etadiam \cs(\etadiam) = -1.
\label{eq:etadiam}
\end{equation}
In order to find the transformation from $\ksnd \tau(\etasnd)=-1$ to
\cref{eq:etadiam}, the explicit dependence of $\tau(\eta)$ needs
to be determined. In analogy with the expansion presented in
\cref{sec:hubbleflow}, this one can be perturbatively obtained by
successively integrating \cref{eq:taudef} by parts over the
scale factor $a$. One gets
\begin{equation}
\begin{aligned}
  -\tau & = \int_a^{+\infty} \dfrac{\cs(a)}{a^2 H} \dd{a} \\
  & = \dfrac{\cs}{aH} + \int_a^{+\infty}\left[ \dfrac{1}{a H} \dv{\cs}{a} +
  \dfrac{\cs}{a} \dv{\left(H^{-1}\right)}{a} \right]\dd{a}.
\end{aligned}
\label{eq:tauparts}
\end{equation}
From the definition of the sound flow functions in
\cref{eq:soundflow}, one has
\begin{equation}
\dv{\cs}{a} = -\dfrac{\cs}{a} \del{1},\qquad \dv{\del{i}}{a} =
\dfrac{\del{i} \del{i+1}}{a}\,.
\end{equation}
Replacing all derivative terms in \cref{eq:tauparts} by the sound
and Hubble flow functions, and successively integrating by parts, one
obtains
\begin{equation}
\begin{aligned}
  -\tau & = \dfrac{\cs}{\calH} \left(1 + \eps{1} - \del{1} + \eps{1}^2 +
\eps{1}\eps{2} - 2\eps{1} \del{1} + \del{1}^2 - \del{1} \del{2}
\right. \\ & \left. +
\eps{1}^3 + 3 \eps{1}^2 \eps{2} + \eps{1} \eps{2}^2 + \eps{1} \eps{2}
\eps{3} - 3\eps{1}^2 \del{1} - 3 \eps{1} \eps{2} \del{1} \right. \\ & \left. + 3 \del{1}^2
\eps{1} - 3 \eps{1} \del{1} \del{2} - \del{1}^3 + 3 \del{1}^2 \del{2}
- \del{1} \del{2}^2 - \del{1} \del{2} \del{3} \right) \\& + \order{\eps{}^4}.
\label{eq:tau}
\end{aligned}
\end{equation}
Combined with the expansion of the conformal Hubble parameter $\calH$
given in \cref{eq:eta3rd}, we finally get
\begin{equation}
\begin{aligned}
  \dfrac{\eta \cs}{\tau} & = 1 + \del{1} + \eps{1} \del{1} +
\del{1}\del{2} + 2 \eps{1} \eps{2} \del{1} + \eps{1}^2 \del{1} + 2
\eps{1} \del{1}\del{2} \\ & + \del{1}\del{2}\del{3} + \del{1} \del{2}^2 -
\del{1}^2 \del{2} + \order{\eps{}^4}.
\end{aligned}
\label{eq:etacsotausr3}
\end{equation}

Performing the change of pivot can be made in two strictly equivalent
ways. Either implicitly assuming that $\ksnd = \kdiam$ and
re-expressing all the flow functions $\epssnd{i}$ and $\delsnd{i}$ in
terms of $\epsdiam{i}$ and $\deldiam{i}$ by their expansion in
$\ln(\etasnd/\etadiam)$, or, implicitly assuming that
$\etasnd=\etadiam$ and remarking that \cref{eq:etasnd} divided
by \cref{eq:etadiam} gives
\begin{equation}
  \ln(\ksnd) = \ln(\kdiam) + \ln\left[\dfrac{\etadiam \cs(\etadiam)}{\tau(\etasnd)} \right].
\end{equation}
Choosing the latter way, plugging $\etadiam=\etasnd$ into the previous
equation, we get, with the help of \cref{eq:etacsotausr3}, the
simple relation
\begin{equation}
\begin{aligned}
  \ln\left(\dfrac{k}{\ksnd}\right) & =  \ln\left(\dfrac{k}{\kdiam}\right) - \deldiam{1} -
\deldiam{1}\epsdiam{1} - \deldiam{1} \deldiam{2} +
\dfrac{1}{2}\deldiam{1}^2  \\ &- 2\deldiam{1} \epsdiam{1} \epsdiam{2} -
\deldiam{1} \epsdiam{1}^2 - 2 \deldiam{1} \deldiam{2} \epsdiam{1} +
\deldiam{1}^2 \epsdiam{1} \\ & - \deldiam{1} \deldiam{2} \deldiam{3} -
\deldiam{1} \deldiam{2}^2 + 2 \deldiam{1}^2 \deldiam{2} - \dfrac{1}{3}
\deldiam{1}^3 + \order{\eps{}^4}.
\end{aligned}
\end{equation}
\begin{widetext}
\subsubsection{Power spectrum}

Using this formula in \cref{eq:calPzsr3snd}, one gets the power
spectrum of the curvature perturbation for K-inflation, at N3LO, fully
expanded around the pivot wavenumber $\kdiam$,
\begin{equation}
\begin{aligned}
  \calPz(k) & = \dfrac{\Hdiam^2}{8 \pi^2 \epsdiam{1}
  \csdiam}\left[\bsdiam{0} + \bsdiam{1} \ln
  \left(\dfrac{k}{\kdiam}\right) + \bsdiam{2}
  \ln^2\left(\dfrac{k}{\kdiam}\right)  + \bsdiam{3}
  \ln^3\left(\dfrac{k}{\kdiam}\right) \right].
\end{aligned}
\label{eq:calPzsr3diam}
\end{equation}
The first term encodes the corrections to the amplitude and reads
\begin{equation}
\begin{aligned}
\bsdiam{0} & =
    1
    - 2 \qty(C + 1) \epsdiam{1}
    - C \epsdiam{2}
    + \qty(C + 2) \deldiam{1}
    + \frac{1}{2} (\pi^2 + 4 C^2 + 4 C - 6) \epsdiam{1}^2
    + \frac{1}{12} (7 \pi^2 + 12 C^2 - 12 C - 72) \epsdiam{1} \epsdiam{2} \\
    & - \frac{1}{2} (\pi^2 + 4 C^2 + 6 C - 8) \epsdiam{1} \deldiam{1}
    + \frac{1}{8} (\pi^2 + 4 C^2 - 8) \epsdiam{2}^2
    + \frac{1}{24} (\pi^2 - 12 C^2) \epsdiam{2} \epsdiam{3}
    - \frac{1}{4} (\pi^2 + 4 C^2 + 4 C - 12) \epsdiam{2} \deldiam{1} \\
    & + \frac{1}{8} (\pi^2 + 4 C^2 + 8 C - 8) \deldiam{1}^2
    - \frac{1}{24} (\pi^2 - 12 C^2 - 48 C - 48) \deldiam{1} \deldiam{2}
    \\ & - \frac{1}{24} \qty[4 C^3 + (3 \pi^2 - 8) C + 14 \zeta(3) - 16] (8 \epsdiam{1}^3 + \epsdiam{2}^3) \\
    & + \frac{1}{12} \qty[13 \pi^2 - 8 (\pi^2 - 9) C + 36 C^2 - 84 \zeta(3)] \epsdiam{1}^2 \epsdiam{2}
    + \frac{1}{8} \qty[4 C^3 + (3 \pi^2 - 32) C + 14 \zeta(3) - 16] (4 \epsdiam{1}^2 \deldiam{1} + \epsdiam{2}^2 \deldiam{1}) \\
    & - \frac{1}{24} \qty[8 C^3 - 15 \pi^2 + 6 (\pi^2 - 4) C - 12 C^2 + 100 \zeta(3) + 16] \epsdiam{1} \epsdiam{2}^2 \\
    & + \frac{1}{12} \qty[8 C^3 + \pi^2 + 6 (\pi^2 - 12) C - 12 C^2 - 8 \zeta(3) - 8] \epsdiam{1} \epsdiam{2} \epsdiam{3} \\
    & + \frac{1}{24} \qty[24 C^3 - 13 \pi^2 + 2 (13 \pi^2 - 132) C - 36 C^2 + 168 \zeta(3) - 48] \epsdiam{1} \epsdiam{2} \deldiam{1} \\
    & - \frac{1}{8} \qty[4 C^3 + (3 \pi^2 - 40) C + 14 \zeta(3) - 20] (2 \epsdiam{1} \deldiam{1}^2 + \epsdiam{2} \deldiam{1}^2) \\
    & - \frac{1}{24} \qty[12 C^3 + 13 \pi^2 + (5 \pi^2 - 24) C + 36 C^2 - 120] (2 \epsdiam{1} \deldiam{1} \deldiam{2} + \epsdiam{2} \deldiam{1} \deldiam{2})
    + \frac{1}{24} \qty[12 C^3 + (5 \pi^2 - 48) C] \epsdiam{2}^2 \epsdiam{3} \\
    & + \frac{1}{24} \qty[\pi^2 C - 4 C^3 - 8 \zeta(3) + 16] (\epsdiam{2} \epsdiam{3}^2 + \epsdiam{2} \epsdiam{3} \epsdiam{4})
    - \frac{1}{24} \qty[12 C^3 + (5 \pi^2 - 72) C] \epsdiam{2} \epsdiam{3} \deldiam{1} \\
    & + \frac{1}{24} \qty[4 C^3 + 3 (\pi^2 - 16) C + 14 \zeta(3) - 28] \deldiam{1}^3
    + \frac{1}{24} \qty[12 C^3 + 13 \pi^2 + (5 \pi^2 - 48) C + 36 C^2 - 168] \deldiam{1}^2 \deldiam{2} \\
    & + \frac{1}{24} \qty[4 C^3 - 2 \pi^2 - (\pi^2 - 48) C + 24 C^2 + 8 \zeta(3) + 32] (\deldiam{1} \deldiam{2}^2 + \deldiam{1} \deldiam{2} \deldiam{3}).
\end{aligned}
\label{eq:bsdiam_0}
\end{equation}

The second term gives the deviations from scale invariance and reads
\begin{equation}
\begin{aligned}
\bsdiam{1} & =
    - 2 \epsdiam{1} - \epsdiam{2} + \deldiam{1}
    + (2 C + 1) (2 \epsdiam{1}^2 - \epsdiam{2} \deldiam{1})
    + (2 C - 1) \epsdiam{1} \epsdiam{2}
    - (4 C + 3) \epsdiam{1} \deldiam{1}
    + C (\epsdiam{2}^2 - \epsdiam{2} \epsdiam{3})
    + (C + 1) \deldiam{1}^2 \\
    & + (C + 2) \deldiam{1} \deldiam{2}
    - \frac{1}{8} (\pi^2 + 4 C^2 - 8) (8 \epsdiam{1}^3 + \epsdiam{2}^3)
    - \frac{2}{3} (\pi^2 - 9 C - 9) \epsdiam{1}^2 \epsdiam{2}
    + \frac{1}{8} (3 \pi^2 + 12 C^2 - 32) (4 \epsdiam{1}^2 \deldiam{1} + \epsdiam{2}^2 \deldiam{1}) \\
    & - \frac{1}{4} (\pi^2 + 4 C^2 - 4 C - 4) \epsdiam{1} \epsdiam{2}^2
    + \frac{1}{2} (\pi^2 + 4 C^2 - 4 C - 12) \epsdiam{1} \epsdiam{2} \epsdiam{3}
    + \frac{1}{12} (13 \pi^2 + 36 C^2 - 36 C - 132) \epsdiam{1} \epsdiam{2} \deldiam{1} \\
    & - \frac{1}{8} (3 \pi^2 + 12 C^2 - 40) (2 \epsdiam{1} \deldiam{1}^2 + \epsdiam{2} \deldiam{1}^2)
    - \frac{1}{24} (5 \pi^2 + 36 C^2 + 72 C - 24) (2 \epsdiam{1} \deldiam{1} \deldiam{2} + \epsdiam{2} \deldiam{1} \deldiam{2}) \\
    & + \frac{1}{24} (5 \pi^2 + 36 C^2 - 48) \epsdiam{2}^2 \epsdiam{3}
    + \frac{1}{24} (\pi^2 - 12 C^2) (\epsdiam{2} \epsdiam{3}^2 + \epsdiam{2} \epsdiam{3} \epsdiam{4})
    -\frac{1}{24} (5 \pi^2 + 36 C^2 - 72) \epsdiam{2} \epsdiam{3} \deldiam{1} \\
    & + \frac{1}{8} (\pi^2 + 4 C^2 - 16) \deldiam{1}^3
    + \frac{1}{24} (5 \pi^2 + 36 C^2 + 72 C - 48) \deldiam{1}^2 \deldiam{2}
    - \frac{1}{24} (\pi^2 - 12 C^2 - 48 C - 48) (\deldiam{1} \deldiam{2}^2 + \deldiam{1} \deldiam{2} \deldiam{3}).
  \end{aligned}
\label{eq:bsdiam_1}
\end{equation}
The higher order terms encode deviations from a pure power law and are
\begin{equation}
\begin{aligned}
\bsdiam{2} & =
    2 \epsdiam{1}^2
    + \epsdiam{1} \epsdiam{2}
    - 2 \epsdiam{1} \deldiam{1}
    + \frac{1}{2} \epsdiam{2}^2
    - \frac{1}{2} \epsdiam{2} \epsdiam{3}
    -\epsdiam{2} \deldiam{1}
    + \frac{1}{2} \deldiam{1}^2
    + \frac{1}{2} \deldiam{1} \deldiam{2}
    - 4 C \epsdiam{1}^3
    + 3 \epsdiam{1}^2 \epsdiam{2}
    + 6 C \epsdiam{1}^2 \deldiam{1} \\
    & + \frac{1}{2} (2 C - 1) (2 \epsdiam{1} \epsdiam{2} \epsdiam{3} - \epsdiam{1} \epsdiam{2}^2  + 3 \epsdiam{1} \epsdiam{2} \deldiam{1})
    - 3 C \epsdiam{1} \deldiam{1}^2
    - 3 (C + 1) \epsdiam{1} \deldiam{1} \deldiam{2}
    - \frac{C}{2} (\epsdiam{2}^3 + \epsdiam{2} \epsdiam{3}^2 + \epsdiam{2} \epsdiam{3} \epsdiam{4} - \deldiam{1}^3)\\
    & + \frac{3}{2} C (\epsdiam{2}^2 \epsdiam{3} + \epsdiam{2}^2 \deldiam{1} - \epsdiam{2} \epsdiam{3} \deldiam{1} - \epsdiam{2} \deldiam{1}^2)
    - \frac{3}{2} (C + 1) (\epsdiam{2} \deldiam{1} \deldiam{2} - \deldiam{1}^2 \deldiam{2})
    + \frac{1}{2} (C + 2) (\deldiam{1} \deldiam{2}^2 + \deldiam{1} \deldiam{2} \deldiam{3}),
\end{aligned}
\label{eq:bsdiam_2}
\end{equation}
and
\begin{equation}
\begin{aligned}
\bsdiam{3} & =
    - \frac{4}{3} \epsdiam{1}^3
    + 2 \epsdiam{1}^2 \deldiam{1}
    - \frac{1}{3} \epsdiam{1} \epsdiam{2}^2
    + \frac{2}{3} \epsdiam{1} \epsdiam{2} \epsdiam{3}
    + \epsdiam{1} \epsdiam{2} \deldiam{1}
    - \epsdiam{1} \deldiam{1}^2
    - \epsdiam{1} \deldiam{1} \deldiam{2}
    - \frac{1}{6} \epsdiam{2}^3
    + \frac{1}{2} \epsdiam{2}^2 \epsdiam{3}
    + \frac{1}{2} \epsdiam{2}^2 \deldiam{1} \\
    & - \frac{1}{6} \epsdiam{2} \epsdiam{3}^2
    - \frac{1}{6} \epsdiam{2} \epsdiam{3} \epsdiam{4}
    - \frac{1}{2} \epsdiam{2} \epsdiam{3} \deldiam{1}
    - \frac{1}{2} \epsdiam{2} \deldiam{1}^2
    - \frac{1}{2} \epsdiam{2} \deldiam{1} \deldiam{2}
    + \frac{1}{6} \deldiam{1}^3
    + \frac{1}{2} \deldiam{1}^2 \deldiam{2}
    + \frac{1}{6} \deldiam{1} \deldiam{2}^2
    + \frac{1}{6} \deldiam{1} \deldiam{2} \deldiam{3}.
  \end{aligned}
\label{eq:bsdiam_3}
\end{equation}
\end{widetext}

\subsubsection{Tensor modes}

During K-inflation, the tensor modes evolution remains unaffected by
any modification in the scalar sector, and, as such, their primordial
power spectrum is still given by \cref{eq:calPhsr3ast}. However,
we still need to re-express \cref{eq:calPhsr3ast} in terms of the
Hubble flow functions evaluated at the same time as the ones appearing
in the scalar power spectrum, namely at $\etadiam$. As discussed in
\cref{sec:pivot}, the easiest way to implement the change of pivot
is to implicitly identify $\etaast$ and $\etadiam$ while letting the
pivot wavenumber running. From \cref{eq:etaast,eq:etadiam}, we have
\begin{equation}
\ln\left(\kast\right) = \ln\left(\kdiam\right) +
\ln\left(\dfrac{\etadiam \csdiam}{\etaast}\right),
\label{eq:kast2kdiam}
\end{equation}
such that the tensor power spectrum $\calPh(k)$, at N3LO, fully
expanded around the pivot $\etadiam$ is given by
\cref{eq:calPhsr3ast} with the following simple replacement rule
\begin{equation}
\kast \to \kdiam \csdiam, \qquad \epsast{i} \to \epsdiam{i}.
\label{eq:ast2diam}
\end{equation}
Because this transformation is trivial, and for the sake of clarity,
we do not rewrite the full expression of $\calPh(k)$ at
$\etadiam$. Let us notice that, if one expands all the logarithms
$\ln[k/(\kdiam \csdiam)]$, the only formal difference with respect to
\cref{eq:calPhsr3ast} are terms which are powers of $\ln(\cs)$. Up
to N2LO, these terms match the ones previously derived in
Ref.~\cite{Jimenez:2013xwa}.

\subsubsection{Other possible pivot}

Another way to get consistent formulas for both the scalar and tensor
mode power spectra for K-inflation would be to shift the scalar mode
pivot from $\etadiam$ to $\etaast$. Using \cref{eq:kast2kdiam} again,
we have now the replacement rule
\begin{equation}
\kdiam \to \dfrac{\kast}{\csast}, \qquad \epsdiam{i} \to \epsast{i},
\qquad \deldiam{i} \to \delast{i},
\end{equation}
to be applied to \crefrange{eq:calPzsr3diam}{eq:bsdiam_3}
while the expression for the tensor modes, \cref{eq:calPhsr3ast},
is left unchanged.

\section{Conclusion}
\label{sec:conclusion}
The main results of this work are the explicit expressions of the
tensor and scalar primordial power spectra generated during single
field inflation with a minimal kinetic term, see
\cref{eq:calPhsr3ast,eq:calPzsr3ast}, and, with a non-minimal kinetic
term, see \cref{eq:calPzsr3diam}. Let us notice that, as expected,
forcing $\csdiam=1$ and $\deldiam{i}=0$ into \cref{eq:calPzsr3diam}
gives back \cref{eq:calPzsr3ast}. From a data analysis point of view,
these expressions are the ones that should be used when comparing the
observable predictions of slow-roll inflation to any cosmological
data. As motivated in the introduction, one should expect the next
generation of cosmological observations to provide information on
$\epsdiam{3}$. This slow-roll parameter appears at N2LO and the
expressions at N3LO derived in this paper therefore encode the
dominant theoretical uncertainties. In other words, in order to search
for a non-vanishing running of the spectral index in the data, one
should perform a data analysis involving $\epsdiam{4}$ and marginalize
over it to get a robust posterior probability distribution on the
slow-roll parameter $\epsdiam{3}$.

Another possible interest of the formulas derived here, and more
particularly the ones in \cref{sec:pln4lo}, are for extrapolations of
the power spectra. For instance, if one needs to estimate the
amplitude of the curvature perturbations, or gravitational waves, at
wavenumbers significantly different than $\kast$ (or $\kdiam$), all
higher order terms may play a significant role.

Finally, a last but not least interest of having at our disposal
highly accurate formulas for the semi-classical slow-roll predictions
is to allow for searching in the data unexpected deviations. As
mentioned in the introduction, for standard slow-roll, quantum
backreaction is expected to be negligible, genuinely not showing up
before the sixth order in the slow-roll parameters. However, the
presence in the potential of any flat regions, or inflection points,
between the time at which observable modes left the Hubble radius and
the end of inflation are expected to boost quantum
backreaction~\cite{Ando:2020fjm}. As such, it could induce small
deviations to the semi-classical slow-roll predictions, the precise
form of which would still need to be determined.

\begin{acknowledgments}
It is a pleasure to thank Vincent Vennin for enlightening discussions
as well as an anonymous referee for their careful reading of the
manuscript. This work is supported by the ``Fonds de la Recherche
Scientifique - FNRS'' under Grant $\mathrm{N^{\circ}T}.0198.19$ as
well as by the Wallonia-Brussels Federation Grant ARC
$\mathrm{N^{\circ}}19/24-103$.

\end{acknowledgments}

\appendix

\section{Generating functionals}
\label{sec:genfunc}

\subsection{The $\F{n}(x)$ hierarchy}
\label{sec:F1d}

To determine the asymptotic expansion of the functions $\F{n}$ defined in \cref{eq:Fn}, let us define the generating functional
\begin{equation}
    \Fg(\nu, x) \equiv \sum_n \frac{\nu^n}{n!} \F{n}(x),
\end{equation}
so that the functions $\F{n}$ satisfy
\begin{equation}
    \F{n}(x) = \eval{\pdv[n]{\Fg(\nu, x)}{\nu}}_{\nu=0}.
\end{equation}
Hence, the functional $f(\nu,x)$ completely determines the behavior of the functions $\F{n}$.

While it is difficult to determine the functions $\F{n}(x)$ for any $n$, it is easy to find the form of the functional $f(\nu, x)$.
Indeed
\begin{align}
    f(\nu, x) &= \int_x^\infty \qty(\sum_n \frac{\nu^n \ln^n u}{n!}) \frac{e^{2iu}}{u} \dd{u} \\
    &= \int_x^\infty u^{\nu - 1} e^{2iu}\dd{u} = x^\nu \mathrm{E}_{1-\nu} (-2ix)
\end{align}
in which $\mathrm{E}_{1-\nu}$ is the generalized exponential integral.

In this paper, we only need the asymptotic expansion of $\F{n}(x)$ in the limit $x \to 0$.
Therefore, we only need an asymptotic form of $f(\nu, x)$ in this limit
\begin{equation}
    f(\nu, x) \underset{x \to 0}{\sim} -\frac{x^{\nu }}{\nu }+2^{-\nu } e^{i \pi  \nu / 2} \Gamma (\nu ).
\end{equation}
From this expression, we obtain systematically the asymptotic behavior of all the functions $\F{n}$
\begin{align}
    F_0(x) &= - \ln x - B + \order{x},\\
    F_1(x) &= - \frac{1}{2}\ln^2x + \frac{B^2}{2} + \frac{\pi^2}{12} + \order{x}, \\
    F_2(x) &=  - \frac{1}{3}\ln ^3 x - \frac{B^3}{3} - \frac{\pi^2}{6}B - \frac{2}{3}\zeta(3)  + \order{x},
\end{align}
where we have used the complex constant
\begin{equation}
B \equiv \gammaE + \ln(2) - \dfrac{i \pi}{2}\,,
  \label{eq:Bdef}
\end{equation}
and with $\gammaE$ the Euler-Mascheroni constant.

\subsection{The $\F{0^n}$ hierarchy}
\label{sec:Fnd}
Let us introduce the hierarchy of integrals $\I{n}$ defined by
\begin{equation}
      \I{n+1} = \int_x^{+\infty} \dfrac{e^{+2iy}}{y} \conj{\I{n}}(y)
      \dd{y}, \qquad \I{0} = 1.
\label{eq:Indef}
\end{equation}
From this definition, we see that
\begin{equation}
\I{2n}(x) = \conj{\F{0^{2n}}}(x), \qquad \I{2n+1}(x) = \F{0^{2n+1}}(x),
\label{eq:I2F}
\end{equation}
while the $\I{n}(x)$ are solutions of an infinite coupled system of
complex differential equations
\begin{equation}
\dv{\I{n+1}}{x} = - \dfrac{e^{+2ix}}{x} \conj{\I{n}}(x).
\end{equation}
A generating functional $\Ig(\nu,x)$ can be constructed as
\begin{equation}
\Ig(\nu,x) \equiv \sum_{k=0}^{+\infty} \I{k}(x) \nu^k,
\label{eq:Igdef}
\end{equation}
from which one has
\begin{equation}
\I{n}(x) = \dfrac{1}{n!}\eval{\pdv[n]{\Ig(\nu,x)}{\nu}}_{\nu=0}.
\label{eq:Igtaylor}
\end{equation}
From \cref{eq:Indef}, the generating functional verifies the
complex differential equation
\begin{equation}
\pdv{\Ig}{x} + \dfrac{\nu e^{2ix} }{x}  \conj{\Ig}(x) = 0.
\end{equation}
Separating the real and imaginary parts with $\Ig = a(x) + i b(x)$,
one obtains the differential system
\begin{equation}
  \begin{aligned}
    \dv{a}{x} & = -\dfrac{\nu}{x} \cos(2x) a - \dfrac{\nu}{x} \sin(2x) b, \\
    \dv{b}{x} & = -\dfrac{\nu}{x} \sin(2x) a + \dfrac{\nu}{x} \cos(2x) b.
  \end{aligned}
\end{equation}
Defining
\begin{equation}
  X(x) \equiv
  \begin{bmatrix}
    a(x) \\
    b(x)
  \end{bmatrix}
  ,\qquad
  A(x) \equiv
  \begin{bmatrix}
    \cos(2x) & \sin(2x) \\ \sin(2x) & -\cos(2x)
  \end{bmatrix},
\end{equation}
it can be recast into the matrix form
\begin{equation}
\dv{X}{x} = -\dfrac{\nu}{x} A X.
\label{eq:dXodx}
\end{equation}
The matrix $A$ can be diagonalized by a rotation as
\begin{equation}
A = P \Lambda P^{-1},
\end{equation}
where
\begin{equation}
  \Lambda =
  \begin{pmatrix}
    1 & 0 \\
    0 & -1
  \end{pmatrix},
  \qquad
  P =
  \begin{bmatrix}
    \cos(x) & -\sin(x) \\
    \sin(x) & \cos(x)
  \end{bmatrix}.
\end{equation}
We can now perform the change of variable
\begin{equation}
  Z(x) =
  \begin{bmatrix}
    z_1(x) \\
    z_2(x)
  \end{bmatrix}
  \equiv
    P^{-1} X,
\label{eq:zdef}
\end{equation}
in terms of which \cref{eq:dXodx} greatly simplifies into
\begin{equation}
\begin{aligned}
  \dv{Z}{x} & = -\left(\dfrac{\nu}{x} \Lambda + P^{-1}\dv{P}{x} \right) Z
  & =
  \begin{pmatrix}
    -\dfrac{\nu}{x} & 1 \\
    -1 & \dfrac{\nu}{x}
  \end{pmatrix} Z .
\end{aligned}
\end{equation}
This system has no longer any oscillatory terms and can be decoupled
by differentiation. One gets
\begin{equation}
  \begin{aligned}
  \dv[2]{z_1}{x}  & = -  \left[1- \dfrac{\nu(\nu+1)}{x^2} \right] z_1, \qquad
  z_2   = \dv{z_1}{x} + \dfrac{\nu}{x} z_1.
  \end{aligned}
\label{eq:magicattibessel}
\end{equation}
The first of these equations is, once more, a Riccati-Bessel
differential equation which admits the exact solutions
\begin{equation}
z_1(x) = C_1(\nu) \, x \sbesselj{\nu}(x) + C_2(\nu) \, x \sbessely{\nu}(x),
\label{eq:z1sol}
\end{equation}
where $\sbesselj{\nu}$ and $\sbessely{\nu}$ are the spherical Bessel
functions of first and second kind~\cite{Abramovitz:1970aa}.
The yet undetermined functions $C_1(\nu)$ and $C_2(\nu)$ are the
integration constants with respect to $x$. Plugging \cref{eq:z1sol}
into \cref{eq:magicattibessel}, and using the standard recurrence
relations on the spherical Bessel functions, one obtains
\begin{equation}
z_2(x) = C_1(\nu) \,  x \sbesselj{\nu-1}(x) + C_2(\nu) \, x \sbessely{\nu-1}(x).
\label{eq:z2sol}
\end{equation}
From the definition of the $\I{n}(x)$ in \cref{eq:Indef},
assuming a smooth convergence at infinity and no branch cut, we have,
for all $\nu$
\begin{equation}
\lim_{x\to \infty} \Ig(\nu,x) = 1.
\end{equation}
This condition fixes the integration constants $C_1(\nu)$ and
$C_2(\nu)$, which, by using the asymptotic forms of the spherical
Bessel functions, read
\begin{equation}
C_1(\nu) = -\sin\left(\dfrac{\pi \nu}{2}\right), \qquad C_2(\nu) =
-\cos\left(\dfrac{\pi \nu}{2}\right).
\end{equation}
From \cref{eq:zdef,eq:z1sol,eq:z2sol}, one
finally obtains the exact expression for the generating functional
\begin{equation}
\begin{aligned}
  \Ig(\nu,x) & =  -x e^{ix} \left\{ \sin\left(\dfrac{\pi \nu}{2}
  \right) \left[\sbesselj{\nu}(x) + i \sbesselj{\nu-1}(x) \right]
  \right. \\ & \left. +
  \cos\left(\dfrac{\pi \nu}{2}\right) \left[\sbessely{\nu}(x) + i
    \sbessely{\nu-1}(x) \right] \right\}.
\end{aligned}
\label{eq:Igexact}
\end{equation}
The exact expression of all the integrals $\F{0^n}(x)$ can now be
obtained from \cref{eq:I2F,eq:Igtaylor,eq:Igexact}, i.e., by
differentiation of the spherical Bessel functions with respect to
$\nu$. Let us mention that some of these integrals have been discussed
in Ref.~\cite{Milgram1985}.

For our purpose, the $\F{0^n}(x)$ need to be evaluated only
in the super-Hubble limit $x\to0$, and, in practice, we only need an
asymptotic form of $\Ig(\nu,x)$ in the same limit. From
\cref{eq:Igexact}, using the expansion of the spherical Bessel
functions near the origin~\cite{Abramovitz:1970aa}, we obtain
\begin{equation}
\begin{aligned}
  \Ig(\nu,x) & \underset{x \to 0}{\sim} \dfrac{2^\nu}{x^\nu
    \sqrt{\pi}} \cos\left(\dfrac{\pi \nu}{2}\right) \Gamma\left(\nu +
  \dfrac{1}{2}\right) \\ & + \dfrac{x^\nu \sqrt{\pi}}{2^\nu \cos(\pi
    \nu)} \dfrac{i \sin\left(\dfrac{\pi \nu}{2}
    \right)}{\Gamma\left(\nu + \dfrac{1}{2}\right)}.
\end{aligned}
\label{eq:Igapprox}
\end{equation}
From \cref{eq:Igtaylor}, we have near the origin
\begin{equation}
\F{0}(x) = -B - \ln(x) + \order{x},
\end{equation}
where $B$ is given in \cref{eq:Bdef}. This expression matches the
one derived by the other method of \cref{sec:F1d}. The second
derivative of $\Ig(\nu,x)$ with respect to $\nu$ gives
\begin{equation}
\F{00}(x) = \frac{\pi^2}{4} + \frac{B^2}{2} + B \ln(x) +
\frac{1}{2} \ln^2(x) + \order{x},
\end{equation}
while the third derivative allows us to determine
\begin{equation}
  \begin{aligned}
\F{000}(x) & = - \frac{7}{3}  \zeta(3) -\frac{\pi^2}{4} B - \frac{1}{6}
B^3 - \left(\frac{\pi^2}{4} +  \frac{B^2}{2}\right) \ln(x) \\ &
- \frac{B}{2} \ln^2(x) - \frac{1}{6} \ln^3(x) + \order{x}.
  \end{aligned}
\end{equation}

\begin{widetext}
\section{Direct calculation}

\label{sec:details}
\subsection{General solution}
\label{sec:gdetails}
As an intermediate step, using \cref{eq:mu0,eq:mu1,eq:mu2,eq:ahat}, we give the super-Hubble limit for the
numerator. At N3LO in slow-roll, one gets
\begin{equation}
\begin{aligned}
	\left|x \muh{} \right|^2 &= 1
 	+ \frac{2}{3} \gk{1} \qty[2 + \Re(\F{0})]
	+ \frac{2}{27} \gk{1}^2 \qty[4 + 3 \abs{\F{0}}^2 + 11 \Re(\F{0})]
	+ \frac{2}{9} \gk{2} \qty[8 + 7 \Re(\F{0}) + 3 \Re(\F{1}) + 6 \ln x]
	\\ & + \frac{2}{243} \gk{1}^3 \qty[-8 + 14 \Re(\F{0}) + 30 \abs{\F{0}}^2 + 9 \Re(\overline{\F{00}} \F{0}) + 9 \Re(\F{000})]
	 \\ & + \frac{4}{81} \gk{1}\gk{2} \qty[- 4 + 21 \abs{\F{0}}^2 + 9 \Re(\overline{\F{1}} \F{0}) + 40 \Re(\F{0}) + 15 \Re(\F{1}) + 12 \ln x + 18 \Re(\F{0}) \ln x]
	 \\ & + \frac{2}{27} \gk{3} \qty[52 + 50 \Re(\F{0}) + 42 \Re(\F{1}) + 9 \Re(\F{2}) + 48 \ln x + 18 \ln^2x] + \order{\epsilon^4, x},
\end{aligned}
\label{eq:x2mu2}
\end{equation}
where the $x$-dependency of the integrals has been omitted. Let us
notice we have removed the integrals $\F{01}(x)$ and $\F{10}(x)$ from this
expression since they only appear as the peculiar combination
\begin{equation}
\Re{\left[\F{10}(x) + \overline{\F{01}}(x)\right]} = \Re{\left[\conj{\F{0}}(x)
    \F{1}(x) \right]}.
\label{eq:realmagic}
\end{equation}
This equality stems from an integration by parts of $\F{01}(x)$, given
in \cref{eq:Fnp}, which gives the relation
\begin{equation}
\F{01}(x) + \overline{\F{10}}(x) = \overline{\F{0}}(x) \F{1}(x).
\end{equation}
Note that we have also simplified \cref{eq:x2mu2} using a similar equality
\begin{equation}
\F{00} + \overline{\F{00}} = \F{0} \overline{\F{0}}.
\end{equation}

\subsection{Application to scalars}
\label{sec:sdetails}

In this section, we derive the scalar power spectrum at N3LO in the
pivot of \cref{eq:etasnd}.  For completeness, we use the method of the
Green's function all the way through and this allows us to cross-check
the mapping technique described in
\cref{sec:mapping,sec:mapping-tensor}.  A fearless reader may indeed
notice that the two methods yield identical results, as expected.

From the definition of the new time $\tau$ in \cref{eq:taudef}, one finds
\begin{equation}
    \dv{\tau} = \frac{1}{\cs} \dv{\eta} = \frac{\calH}{\cs} \dv{N}\,,
\end{equation}
where \cref{eq:Nprime} has been used to express the derivative with
respect to $\eta$.

To use the Green's function method, we need to recast the equation of
motion~\eqref{eq:nonminimal-eom} into the form of \cref{eq:musplit}.
We do so by expressing $z^{-1} \dv*[2]{z}{\tau}$ in terms of
Hubble flow and sound flow functions
\begin{equation}
\begin{aligned}
    \frac{1}{z} \dv[2]{z}{\tau} &=
    \qty(\frac{\calH}{\cs}) ^2 \bigg(
        2
        - \eps1
        + \frac{3}{2} \eps2
        + \frac{5}{2} \del1
        - \frac{1}{2} \eps1 \eps2
        - \frac{1}{2} \eps1 \del1
         + \frac{1}{4} \eps2^{2}
        + \frac{1}{2} \eps2 \eps3
        + \eps2 \del1
        + \frac{3}{4} \del1^2
        + \frac{1}{2} \del1 \del2
    \bigg).
\end{aligned}
\label{eq:d2dz2}
\end{equation}
We recall that $z$ is defined in \cref{eq:zkinf}. Note that the equation above is exact.

From \cref{eq:hubbleflow,eq:tau}, one has
\begin{equation}
\begin{aligned}
    \dv{\eps{i}}{\ln \tau} &= \frac{\tau \calH}{\cs} \eps{i}\eps{i+1}
        = - \eps{i}\eps{i+1} \bigg(
            1 + \eps{1} - \del{1}
            + \eps{1}^2 + \eps{1}\eps{2} - 2\eps{1} \del{1} + \del{1}^2 - \del{1} \del{2}
            + \eps{1}^3 + 3 \eps{1}^2 \eps{2} + \eps{1} \eps{2}^2 \\
            &+ \eps{1} \eps{2} \eps{3}
            - 3\eps{1}^2 \del{1} - 3 \eps{1} \eps{2} \del{1} + 3 \del{1}^2 \eps{1}
            - 3 \eps{1} \del{1} \del{2} - \del{1}^3  + 3 \del{1}^2 \del{2}
            - \del{1} \del{2}^2 - \del{1} \del{2} \del{3}\bigg) + \order{\eps{}^6},
\end{aligned}
\end{equation}
and similarly for the sound flow parameters
\begin{equation}
    \dv{\del{i}}{\ln \tau} = \frac{\tau \calH}{\cs} \del{i}\del{i+1}.
\end{equation}

Extending \cref{eq:d2dz2} around the pivot time $\tausnd$, i.e.,
\begin{equation}
\tausnd = -\dfrac{1}{\ksnd}\,,
\end{equation}
one can identify the coefficients $\gsnd{}$ appearing in a decomposition identical to the one
of \cref{eq:musplit}, but, now, in terms of the variable
$x\equiv-k\tau$. The leading term reads
\begin{equation}
\begin{aligned}
    \gsnd{1} &=
        3 \epssnd{1}
        + \frac{3}{2} \epssnd{2}
        - \frac{3}{2} \delsnd{1}
        + 4 \epssnd{1}^2
        + \frac{13}{2} \epssnd{1} \epssnd{2}
        - \frac{11}{2} \epssnd{1} \delsnd{1}
        + \frac{1}{2} \epssnd{2} \epssnd{3}
        + \frac{1}{4} \epssnd{2}^2
        - 2 \epssnd{2} \delsnd{1}
        + \frac{7}{4} \delsnd{1}^2
        - \frac{7}{2} \delsnd{1} \delsnd{2}
        + 5 \epssnd{1}^3 \\
        & + \frac{35}{2} \epssnd{1}^2 \epssnd{2}
        - \frac{23}{2} \epssnd{1}^2 \delsnd{1}
        + \frac{15}{2} \epssnd{1} \epssnd{2}^2
        + 5 \epssnd{1} \epssnd{2} \epssnd{3}
        - 17 \epssnd{1} \epssnd{2} \delsnd{1}
        + \frac{17}{2} \epssnd{1} \delsnd{1}^2
        - 13 \epssnd{1} \delsnd{1} \delsnd{2}
        - \epssnd{2} \epssnd{3} \delsnd{1}
        - \frac{1}{2} \epssnd{2}^2 \delsnd{1} \\
        & + \frac{5}{2} \epssnd{2} \delsnd{1}^2
        - 3 \epssnd{2} \delsnd{1} \delsnd{2}
        -2 \delsnd{1}^3
        + 10 \delsnd{1}^2 \delsnd{2}
        - 4 \delsnd{1} \delsnd{2}^2
        - 4 \delsnd{1} \delsnd{2} \delsnd{3} \\
        & + \frac{1}{2} \bigg(
            6 \epssnd{1} \epssnd{2}
            + 3 \epssnd{2} \epssnd{3}
            - 3 \delsnd{1} \delsnd{2}
            + 22 \epssnd{1}^2 \epssnd{2}
            + 13 \epssnd{1} \epssnd{2}^2
            + 16 \epssnd{1} \epssnd{2} \epssnd{3}
            - 17 \epssnd{1} \epssnd{2} \delsnd{1}
            - 14 \epssnd{1} \delsnd{1} \delsnd{2}
            + \epssnd{2}^2 \epssnd{3}
            + \epssnd{2} \epssnd{3}^2 \\
            & + \epssnd{2} \epssnd{3} \epssnd{4}
            - 7 \epssnd{2} \epssnd{3} \delsnd{1}
            - 4 \epssnd{2} \delsnd{1} \delsnd{2}
            + 10 \delsnd{1}^2 \delsnd{2}
            - 7 \delsnd{1} \delsnd{2}^2
            - 7 \delsnd{1} \delsnd{2} \delsnd{3}
        \bigg) \ln(\frac{k}{\ksnd}) \\
        & + \frac{3}{4} \bigg(
            2 \epssnd{1} \epssnd{2}^2
            + 2 \epssnd{1} \epssnd{2} \epssnd{3}
            + \epssnd{2} \epssnd{3}^2
            + \epssnd{2} \epssnd{3} \epssnd{4}
            - \delsnd{1} \delsnd{2}^2
            - \delsnd{1} \delsnd{2} \delsnd{3}
        \bigg) \ln^2\qty(\frac{k}{\ksnd})
        + \order{\eps{}^4},
\end{aligned}
\label{eq:gsnd1}
\end{equation}
then, for the second order, one obtains
\begin{equation}
\begin{aligned}
    \gsnd{2} &=
        - 3 \epssnd{1} \epssnd{2}
        - \frac{3}{2} \epssnd{2} \epssnd{3}
        + \frac{3}{2} \delsnd{1} \delsnd{2}
        - 11 \epssnd{1}^2 \epssnd{2}
        - \frac{13}{2} \epssnd{1} \epssnd{2}^2
        - 8 \epssnd{1} \epssnd{2} \epssnd{3}
        + \frac{17}{2} \epssnd{1} \epssnd{2} \delsnd{1}
        + 7 \epssnd{1} \delsnd{1} \delsnd{2}
        - \frac{1}{2} \epssnd{2}^2 \epssnd{3} \\
        & - \frac{1}{2} \epssnd{2} \epssnd{3}^2
        - \frac{1}{2} \epssnd{2} \epssnd{3} \epssnd{4}
        + \frac{7}{2} \epssnd{2} \epssnd{3} \delsnd{1}
        + 2 \epssnd{2} \delsnd{1} \delsnd{2}
        - 5 \delsnd{1}^2 \delsnd{2}
        + \frac{7}{2} \delsnd{1} \delsnd{2}^2
        + \frac{7}{2} \delsnd{1} \delsnd{2} \delsnd{3} \\
        & + \frac{3}{2} (
            - 2 \epssnd{1} \epssnd{2}^2
            - 2 \epssnd{1} \epssnd{2} \epssnd{3}
            - \epssnd{2} \epssnd{3}^2
            - \epssnd{2} \epssnd{3} \epssnd{4}
            + \delsnd{1} \delsnd{2}^2
            + \delsnd{1} \delsnd{2} \delsnd{3}
        ) \ln\qty(\frac{k}{\ksnd})
        + \order{\eps{}^4},
\end{aligned}
\label{eq:gsnd2}
\end{equation}
and, finally,
\begin{equation}
\begin{aligned}
    \gsnd{3} &=
        \frac{3}{2} \epssnd{1} \epssnd{2}^2
        + \frac{3}{2} \epssnd{1} \epssnd{2} \epssnd{3}
        + \frac{3}{4} \epssnd{2} \epssnd{3}^2
        + \frac{3}{4} \epssnd{2} \epssnd{3} \epssnd{4}
        -\frac{3}{4} \delsnd{1} \delsnd{2}^2
        - \frac{3}{4} \delsnd{1} \delsnd{2} \delsnd{3}
        + \order{\eps{}^4}.
\end{aligned}
\label{eq:gsnd3}
\end{equation}
Notice that, contrary to \cref{eq:gi}, the coefficients $\gsnd{}$
depend explicitly on $k$. This is due to our choice of the particular
time $\tausnd$ at which we have performed the expansion, which is
directly defined from the pivot scale $\ksnd$. On the contrary, the
dependence on $k$ is implicit in \cref{eq:gi} due to the choice of the
$k$-dependent pivot time of \cref{eq:etak}, namely,
\begin{equation}
    \etak = -\dfrac{1}{k}\,.
\end{equation}
This implies that all the $\epsk{}$ are implicit functions of
$k$. Replacing the coefficients $\gsnd{}$ in \cref{eq:x2mu2} by their
expansion given in \cref{eq:gsnd1,eq:gsnd2,eq:gsnd3}, after some
algebra, one finds the scalar power spectrum at the pivot scale
$\ksnd$
\begin{equation}
\begin{aligned}
\calPz(k) =
    \dfrac{\Hsnd^2}{8 \pi^2 \epssnd{1}
    \cssnd}\left[\bssnd{0} + \bssnd{1} \ln
    \left(\dfrac{k}{\ksnd}\right) + \bssnd{2}
    \ln^2\left(\dfrac{k}{\ksnd}\right)  + \bssnd{3}
    \ln^3\left(\dfrac{k}{\ksnd}\right) \right].
\end{aligned}
\label{eq:calPzsr3snd}
\end{equation}
The corrections to the amplitude read
\begin{equation}
\begin{aligned}
\bssnd{0} & =
    1
    - 2 (C + 1) \epssnd{1}
    - C \epssnd{2}
    + (C + 2) \delsnd{1}
    + \frac{1}{2} (\pi^2 + 4 C^2 + 4 C - 6) \epssnd{1}^2
    + \frac{1}{12} (7 \pi^2 + 12 C^2 - 12 C - 72) \epssnd{1} \epssnd{2} \\
    & - \frac{1}{2} (\pi^2 + 4 C^2 + 6 C - 4) \epssnd{1} \delsnd{1}
     + \frac{1}{8} (\pi^2 + 4 C^2 - 8) \epssnd{2}^2
    + \frac{1}{24} (\pi^2 - 12 C^2) \epssnd{2} \epssnd{3}
    - \frac{1}{4} (\pi^2 + 4 C^2 + 4 C - 8) \epssnd{2} \delsnd{1} \\
    & + \frac{1}{8} (\pi^2 + 4 C^2 + 8 C) \delsnd{1}^2
    - \frac{1}{24} (\pi^2 - 12 C^2 - 48 C - 48) \delsnd{1} \delsnd{2} \\
    & + \frac{1}{24} \qty[4 C^3 + 3 (\pi^2 - 8) C + 14 \zeta(3) - 16] (- 8 \epssnd{1}^3 + 12 \epssnd{1}^2 \delsnd{1} - 6 \epssnd{1} \delsnd{1}^2 - \epssnd{2}^3 + 3 \epssnd{2}^2 \delsnd{1} - 3 \epssnd{2} \delsnd{1}^2 + \delsnd{1}^3) \\
    & + \frac{1}{12} \qty[13 \pi^2 - 8 (\pi^2 - 9) C + 36 C^2 - 84 \zeta(3)] \epssnd{1}^2 \epssnd{2} \\
    & - \frac{1}{24} \qty[8 C^3 - 15 \pi^2 + 6 (\pi^2 - 4) C - 12 C^2 + 100 \zeta(3) + 16] \epssnd{1} \epssnd{2}^2 \\
    & + \frac{1}{12} \qty[8 C^3 + \pi^2 + 6 (\pi^2 - 12) C - 12 C^2 - 8 \zeta(3) - 8] \epssnd{1} \epssnd{2} \epssnd{3} \\
    & + \frac{1}{24} \qty[24 C^3 - 13 \pi^2 + 2 (13 \pi^2 - 108) C - 36 C^2 + 168 \zeta(3) - 96] \epssnd{1} \epssnd{2} \delsnd{1} \\
    & + \frac{1}{24} \qty[12 C^3 + 13 \pi^2 + (5 \pi^2 - 24) C + 36 C^2 - 96] (- 2 \epssnd{1} \delsnd{1} \delsnd{2} - \epssnd{2} \delsnd{1} \delsnd{2} + \delsnd{1}^2 \delsnd{2}) \\
    & + \frac{1}{24} \qty[12 C^3 + (5 \pi^2 - 48) C] (\epssnd{2}^2 \epssnd{3} - \epssnd{2} \epssnd{3} \delsnd{1})
    + \frac{1}{24} \qty[\pi^2 C - 4 C^3 - 8 \zeta(3) + 16] (\epssnd{2} \epssnd{3}^2 + \epssnd{2} \epssnd{3} \epssnd{4}) \\
    & + \frac{1}{24} \qty[4 C^3 - 2 \pi^2 - (\pi^2 - 48) C + 24 C^2 + 8 \zeta(3) + 32] (\delsnd{1} \delsnd{2}^2 + \delsnd{1} \delsnd{2} \delsnd{3}).
\end{aligned}
\label{eq:bssnd_0}
\end{equation}
Deviations from scale invariance are encoded in
\begin{equation}
\begin{aligned}
\bssnd{1} & =
    - 2 \epssnd{1}
    - \epssnd{2}
    + \delsnd{1}
    + (2 C + 1) (2 \epssnd{1}^2 - \epssnd{2} \delsnd{1})
    + (2 C - 1) \epssnd{1} \epssnd{2}
    - (4 C + 3) \epssnd{1} \delsnd{1}
    + C (\epssnd{2}^2 - \epssnd{2} \epssnd{3})
    + (C + 1) \delsnd{1}^2 \\
    & + (C + 2) \delsnd{1} \delsnd{2}
    - \frac{1}{8}(\pi^2 + 4 C^2 - 8) (8 \epssnd{1}^3 - 12 \epssnd{1}^2 \delsnd{1} + 6 \epssnd{1} \delsnd{1}^2 + \epssnd{2}^3 - 3 \epssnd{2}^2 \delsnd{1} + 3 \epssnd{2} \delsnd{1}^2 - \delsnd{1}^3) \\
    & - \frac{2}{3} (\pi^2 - 9 C - 9) \epssnd{1}^2 \epssnd{2}
    - \frac{1}{4} (\pi^2 + 4 C^2 - 4 C - 4) \epssnd{1} \epssnd{2}^2
    + \frac{1}{2} (\pi^2 + 4 C^2 - 4 C - 12) \epssnd{1} \epssnd{2} \epssnd{3} \\
    & + \frac{1}{12} (13 \pi^2 + 36 C^2 - 36 C - 108) \epssnd{1} \epssnd{2} \delsnd{1}
    - \frac{1}{24} (5 \pi^2 + 36 C^2 + 72 C - 24) (2 \epssnd{1} \delsnd{1} \delsnd{2} + \epssnd{2} \delsnd{1} \delsnd{2} - \delsnd{1}^2 \delsnd{2}) \\
    & + \frac{1}{24} (5 \pi^2 + 36 C^2 - 48) (\epssnd{2}^2 \epssnd{3} - \epssnd{2} \epssnd{3} \delsnd{1})
    + \frac{1}{24} (\pi^2 - 12 C^2) (\epssnd{2} \epssnd{3}^2 + \epssnd{2} \epssnd{3} \epssnd{4}) \\
    & - \frac{1}{24} (\pi^2 - 12 C^2 - 48 C - 48) (\delsnd{1} \delsnd{2}^2 + \delsnd{1} \delsnd{2} \delsnd{3}).
\end{aligned}
\label{eq:bssnd_1}
\end{equation}
while the deviations from a pure power law spectrum are given by
\begin{equation}
\begin{aligned}
\bssnd{2} & =
    2 \epssnd{1}^2
    + \epssnd{1} \epssnd{2}
    - 2 \epssnd{1} \delsnd{1}
    + \frac{1}{2} \epssnd{2}^2
    - \frac{1}{2} \epssnd{2} \epssnd{3}
    - \epssnd{2} \delsnd{1}
    + \frac{1}{2} \delsnd{1}^2
    + \frac{1}{2} \delsnd{1} \delsnd{2}
    - 4 C \epssnd{1}^3
    + 3 \epssnd{1}^2 \epssnd{2}
    + 6 C \epssnd{1}^2 \delsnd{1} \\
    & - \frac{1}{2} (2 C - 1) (\epssnd{1} \epssnd{2}^2 - 2 \epssnd{1} \epssnd{2} \epssnd{3} - 3 \epssnd{1} \epssnd{2} \delsnd{1})
    - 3 C \epssnd{1} \delsnd{1}^2
    - \frac{3}{2} (C + 1) (2 \epssnd{1} \delsnd{1} \delsnd{2} + \epssnd{2} \delsnd{1} \delsnd{2} - \delsnd{1}^2 \delsnd{2}) \\
    & + \frac{1}{2} C
        (- \epssnd{2}^3 + 3 \epssnd{2}^2 \epssnd{3} + 3 \epssnd{2}^2 \delsnd{1} - \epssnd{2} \epssnd{3}^2 - 3 \epssnd{2} \epssnd{3} \delsnd{1} - \epssnd{2} \epssnd{3} \epssnd{4} - 3 \epssnd{2} \delsnd{1}^2 + \delsnd{1}^3)
    + \frac{1}{2} (C + 2) (\delsnd{1} \delsnd{2}^2 + \delsnd{1} \delsnd{2} \delsnd{3}),
\end{aligned}
\label{eq:bssnd_2}
\end{equation}
and
\begin{equation}
\begin{aligned}
\bssnd{3} & =
    - \frac{4}{3} \epssnd{1}^3
    + 2 \epssnd{1}^2 \delsnd{1}
    - \frac{1}{3} \epssnd{1} \epssnd{2}^2
    + \frac{2}{3} \epssnd{1} \epssnd{2} \epssnd{3}
    - \epssnd{1} \delsnd{1}^2
    - \epssnd{1} \delsnd{1} \delsnd{2}
    + \epssnd{1} \epssnd{2} \delsnd{1}
    - \frac{1}{6} \epssnd{2}^3
    + \frac{1}{2} \epssnd{2}^2 \epssnd{3}
    - \frac{1}{2} \epssnd{2} \delsnd{1} \delsnd{2} \\
    & + \frac{1}{2} \epssnd{2}^2 \delsnd{1}
    - \frac{1}{6} \epssnd{2} \epssnd{3}^2
    - \frac{1}{6} \epssnd{2} \epssnd{3} \epssnd{4}
    - \frac{1}{2} \epssnd{2} \epssnd{3} \delsnd{1}
    - \frac{1}{2} \epssnd{2} \delsnd{1}^2
    + \frac{1}{6} \delsnd{1}^3
    + \frac{1}{2} \delsnd{1}^2 \delsnd{2}
    + \frac{1}{6} \delsnd{1} \delsnd{2}^2
    + \frac{1}{6} \delsnd{1} \delsnd{2} \delsnd{3}.
\end{aligned}
\label{eq:bssnd_3}
\end{equation}

\section{Power-law quantities at N4LO}
\label{sec:pln4lo}

Having at our disposal the expression of the scalar and tensor power
spectra at third order allows us to derive the power law parameters at
N4LO by using the invariance of the results with respect to the choice
of the pivot wavenumber $\kdiam$. The method is detailed in
Ref.~\cite{Schwarz:2004tz} and, here, we apply it using our N3LO
results.

\subsection{Scalar power-law parameters}

The scalar spectral index $\nS$ is defined by
\begin{equation}
\nS - 1 \equiv \eval{\dv{\ln\left[\calPz(k)\right]}{\ln k}}_{k=\kdiam},
\end{equation}
and, to determine its value at N4LO, one can use the invariance of the
power spectrum with respect to the choice of the pivot, i.e.,
\begin{equation}
\dv{\ln \calPz}{\ln\kdiam} = 0.
\label{eq:pivotinv}
\end{equation}
Let us consider the logarithmic expansion
\begin{equation}
\ln\left[\calPz(k)\right] = \ln\Psdiam + \sum_{n=0}^{+\infty}
\dfrac{\lnbsdiam{n}}{n!} \ln^n\left(\dfrac{k}{\kdiam}\right),
\label{eq:lncalPz}
\end{equation}
where the coefficients $\lnbsdiam{n}$ can be straightforwardly
obtained from the $\bsdiam{n}$, i.e., at N3LO in the slow-roll
parameters. In this expression $\Psdiam$ stands for the overall
amplitude
\begin{equation}
  \Psdiam \equiv \dfrac{\Hdiam^2}{8 \pi^2 \epsdiam{1} \csdiam}\,.
\end{equation}
Plugging \cref{eq:lncalPz} into \cref{eq:pivotinv} one gets the
recurrence relations
\begin{equation}
  \begin{aligned}
    \lnbsdiam{1} &= \dv{\lnbsdiam{0}}{\ln\kdiam} + \dv{\ln
      \Psdiam}{\ln\kdiam}, \qquad
    \lnbsdiam{n+1} &= \dv{\lnbsdiam{n}}{\ln\kdiam}.
  \end{aligned}
\end{equation}
Using \cref{eq:etadiam} to trade the derivative with respect to
$\kdiam$ by a derivative with respect to $\etadiam$, one obtains, for
the spectral index,
\begin{equation}
\nS - 1 = \lnbsdiam{1} = \dfrac{1}{\etadiam\calHdiam \deldiam{1}-1}
\left[\dv{\lnbsdiam{0}}{\ln\abs{\etadiam}} + \dv{\ln \Psdiam}{\ln\abs{\etadiam}} \right].
\label{eq:stensm1}
\end{equation}
From the expression of $\eta\calH$ given in
\cref{eq:eta3rd}, and the one of $\bsdiam{0}$ given in \cref{eq:bsdiam_0}, after some basic
but rather long algebra, one gets
\begin{equation}
\nS = 1 + \nsdiam{1} + \nsdiam{2} + \nsdiam{3} + \nsdiam{4} +
\order{\eps{}^5}
\end{equation}
where
\begin{equation}
\begin{aligned}
  \nsdiam{1} & = - 2\epsdiam{1} - \epsdiam{2} +
  \deldiam{1},\\ \nsdiam{2} & = - 2 \epsdiam{1}^2  - \left(2 C + 3\right) \epsdiam{1} \epsdiam{2} - C \epsdiam{2}
  \epsdiam{3} + 3 \epsdiam{1} \deldiam{1}  +
  \epsdiam{2} \deldiam{1} - \deldiam{1}^2 + \left(C +
  2\right) \deldiam{1} \deldiam{2},
\end{aligned}
\label{eq:nsdiam_12}
\end{equation}
and
\begin{equation}
\begin{aligned}
  \nsdiam{3} & = - 2  \epsdiam{1}^3 + \left(\pi^2 - 6  C -
  15\right)\epsdiam{1}^2\epsdiam{2}
  + \frac{1}{12}  \left(7  \pi^2 - 12  C^2 - 36  C
      - 84\right) \epsdiam{1} \epsdiam{2}^2 + \left(
             \frac{7}{12} \pi^2  - C^2 - 4  C - 6 \right)
             \epsdiam{1} \epsdiam{2} \epsdiam{3}
  \\ &  +
    \frac{1}{24}  \left(\pi^2 - 12  C^2\right)
    \epsdiam{2} \epsdiam{3}^2
     + \left(\frac{\pi^2}{4} - 2\right) \epsdiam{2}^2
             \epsdiam{3}
         + \frac{1}{24}  \left(\pi^2 - 12  C^2\right)
              \epsdiam{2} \epsdiam{3}
              \epsdiam{4} -4 \epsdiam{1} \deldiam{1}^2 -\left(\frac{\pi^2}{2} - 4  C - 10 \right)
             \epsdiam{1} \deldiam{1} \deldiam{2}
              \\ & - \left(\frac{\pi^2}{4} - C - 3
             \right)\epsdiam{2} \deldiam{1}\deldiam{2} -\epsdiam{2} \deldiam{1}^2   + 5
              \epsdiam{1}^2 \deldiam{1}  - \left(\frac{\pi^2}{2} - 5
              C - 13\right)\epsdiam{1}\epsdiam{2}\deldiam{1} - \left(
              \frac{\pi^2}{4} - 2  C - 3 \right) \epsdiam{2}
              \epsdiam{3}\deldiam{1}
              \\  & + \deldiam{1}^3  +  \frac{1}{4}  {\left(\pi^2 - 12  C - 32\right)}
    \deldiam{1}^2 \deldiam{2} - \frac{1}{24}
    \left(\pi^2 - 12  C^2 - 48  C - 48\right) \left( \deldiam{1}
    \deldiam{2}^2 + \deldiam{1} \deldiam{2} \deldiam{3}\right).
\end{aligned}
\label{eq:nsdiam_3}
\end{equation}
The \cref{eq:nsdiam_12,eq:nsdiam_3} match the ones
previously derived in Ref.~\cite{Jimenez:2013xwa}, obtained using the
same method but starting from the N2LO corrections to the
amplitude. Our N3LO result allows us to determine the next term, it
reads
\begin{equation}
    \begin{aligned}
      \nsdiam{4} & = - 2 \epsdiam{1}^{4} - \frac{1}{24} {\left[8 C^3 -
          21 \pi^2 - 14 \left(\pi^2 - 12\right) C + 36 C^2 + 100
          \zeta(3) + 88\right]} \epsdiam{1} \epsdiam{2}^3 +
      \left[\frac{1}{2} \left(\pi^2 - 8\right) C \right] \epsdiam{2}^2
      \epsdiam{3}^2 \\ & + \left[\frac{61}{12} \pi^2 + \left(2 \pi^2 -
        35\right) C - 7 C^2 - 14 \zeta(3) - 37\right] \epsdiam{1}^2
      \epsdiam{2}^2 + \left[4 \pi^2 - 12 C - 14 \zeta(3) - 29 \right]
      \epsdiam{1}^3 \epsdiam{2} + \left[2 -\frac{7}{4} \zeta(3)\right]
      \epsdiam{2}^3 \epsdiam{3} \\ & + \left[\frac{17}{6} \pi^2 +
        \left(\pi^2 - 19\right) C - 4 C^2 - 7 \zeta(3) - 20 \right]
      \epsdiam{1}^2 \epsdiam{2} \epsdiam{3} - \left[C^3 - \frac{13}{6}
        \pi^2 - \frac{7}{4} \left(\pi^2 - 12\right) C + 5 C^2 + 9
        \zeta(3) + 9 \right]\epsdiam{1} \epsdiam{2}^2 \epsdiam{3} \\ &
      + \left[-\frac{1}{3} C^3 + \frac{5}{24} \pi^2 + \frac{1}{12}
        \left(7 \pi^2 - 72\right) C - \frac{5}{2} C^2 - \frac{2}{3}
        \zeta(3) - \frac{2}{3}\right] \left(\epsdiam{1} \epsdiam{2}
      \epsdiam{3}^2 + \epsdiam{1} \epsdiam{2} \epsdiam{3} \epsdiam{4}
      \right) + \frac{1}{4} \left(\pi^2 - 8\right) C \epsdiam{2}^2
      \epsdiam{3} \epsdiam{4} \\& + \left[\frac{\pi^2}{8} C -
        \frac{1}{2} C^3 - \zeta(3) + 2\right] \epsdiam{2}
      \epsdiam{3}^2 \epsdiam{4} + \frac{1}{24} {\left[\pi^2 C - 4 C^3
          - 8 \zeta(3) + 16\right]}\left( \epsdiam{2} \epsdiam{3}^3 +
      \epsdiam{2} \epsdiam{3} \epsdiam{4}^2 +\epsdiam{2} \epsdiam{3}
      \epsdiam{4} \epsdiam{5}\right) + 5 \epsdiam{1} \deldiam{1}^3
      \\ & + \left[\frac{13}{4} \pi^2 - 15 C - 7 \zeta(3) -
        44\right]\epsdiam{1} \deldiam{1}^2 \deldiam{2} +
      \left[-\frac{29}{24} \pi^2 - \frac{1}{2} \left(\pi^2 - 24\right)
        C + \frac{5}{2} C^2 + 16\right] \left( \epsdiam{1}
      \deldiam{1}\deldiam{2}^2 + \epsdiam{1} \deldiam{1} \deldiam{2}
      \deldiam{3} \right) \\ & + \epsdiam{2}\deldiam{1}^3 +
      \left[\frac{5}{4} \pi^2 - 3 C - \frac{7}{2} \zeta(3) - 10
        \right] \epsdiam{2} \deldiam{1}^2 \deldiam{2} +
      \left[-\frac{13}{24} \pi^2 - \frac{1}{4} \left(\pi^2 - 12\right)
        C + \frac{1}{2} C^2 + 5\right] \left(\epsdiam{2} \deldiam{1}
      \deldiam{2}^2 + \epsdiam{2} \deldiam{1}\deldiam{2} \deldiam{3}
      \right) \\ & - 9 \epsdiam{1}^2 \deldiam{1}^2 - \left[\frac{5}{2}
        \pi^2 - 9 C - 7 \zeta(3) - 24\right]
      \epsdiam{1}^2\deldiam{1}\deldiam{2} - \left[\frac{\pi^2}{4} -
        \frac{7}{4} \zeta(3)\right] \epsdiam{2}^2
      \deldiam{1}\deldiam{2} + \left[\frac{7}{4} \pi^2 - 9 C -
        \frac{7}{2} \zeta(3) - 28\right] \epsdiam{1} \epsdiam{2}
      \deldiam{1}^2 \\ & - \left[\frac{29}{8} \pi^2 + \left(\pi^2 -
        25\right) C - \frac{9}{2} C^2 - 7 \zeta(3) - 35
        \right]\epsdiam{1} \epsdiam{2} \deldiam{1} \deldiam{2} -
      \left[\frac{5}{8} \pi^2 + \frac{1}{2} \left(\pi^2 - 12\right) C
        - \frac{3}{2} C^2 - 5\right]\epsdiam{2} \epsdiam{3}
      \deldiam{1}\deldiam{2} \\ & + \left[\frac{3}{4} \pi^2 - 3 C -
        \frac{7}{4} \zeta(3) - \frac{13}{2}\right]\epsdiam{2}
      \epsdiam{3} \deldiam{1}^2 + 7 \epsdiam{1}^3 \deldiam{1} -
      \left[\frac{67}{24} \pi^2 + \frac{1}{2} \left(\pi^2 - 32\right)
        C - \frac{7}{2} C^2 - 7 \zeta(3) - 25\right] \epsdiam{1}
      \epsdiam{2}^2 \deldiam{1} \\ &- \left[\frac{\pi^2}{8} +
        \frac{1}{4} \left(\pi^2 - 12\right) C - \frac{3}{2} C^2
        \right] \epsdiam{2} \epsdiam{3}^2 \deldiam{1} -
      \left[\frac{11}{2} \pi^2 - 21 C - 14 \zeta(3) -
        60\right]\epsdiam{2} \epsdiam{1}^2\deldiam{1} \\& -
      \left[\frac{3}{4} \pi^2 - \frac{7}{2} \zeta(3) - 2\right]
      \epsdiam{2}^2 \epsdiam{3} \deldiam{1}- \left[\frac{73}{24} \pi^2
        + \frac{1}{2} \left(\pi^2 - 38\right) C - \frac{7}{2} C^2 - 7
        \zeta(3) - 26\right] \epsdiam{1} \epsdiam{2} \epsdiam{3}
      \deldiam{1} \\ & - \left[\frac{\pi^2}{8} + \frac{1}{4}
        \left(\pi^2 - 12\right) C - \frac{3}{2} C^2 \right]
      \epsdiam{2} \epsdiam{3} \epsdiam{4} \deldiam{1} -
      \deldiam{1}^{4} -\frac{1}{4} {\left[4 \pi^2 - 24 C - 7 \zeta(3)
          - 74\right]} \deldiam{1}^3 \deldiam{2} \\& + \frac{1}{24}
      \left[31 \pi^2 + 12 \left(\pi^2 - 36\right) C - 84 C^2 -
        576\right] \deldiam{1}^2 \deldiam{2}^2\\ & + \frac{1}{24}
      \left[4 C^3 - 2 \pi^2 - \left(\pi^2 - 48\right) C + 24 C^2 + 8
        \zeta(3) + 32\right] \left( \deldiam{1} \deldiam{2}^3 +
      \deldiam{1} \deldiam{2} \deldiam{3}^2 + \deldiam{1} \deldiam{2}
      \deldiam{3} \deldiam{4} \right) \\ & + \left[\frac{2}{3} \pi^2 +
        \frac{1}{4} \left(\pi^2 - 40\right) C - 2 C^2 -
        13\right]\deldiam{1}^2 \deldiam{2} \deldiam{3} +
      \left[\frac{1}{2} C^3 - \frac{1}{4} \pi^2 - \frac{1}{8}
        \left(\pi^2 - 48\right) C + 3 C^2 + \zeta(3) + 4 \right]
      \deldiam{1} \deldiam{2}^2 \deldiam{3}.
    \end{aligned}
    \label{eq:nsm1}
\end{equation}

Using the same technique, we can derive the running of the spectral
index
\begin{equation}
  \alphaS  \equiv \eval{\dv[2]{\ln\left[\calPz(k)\right]}{(\ln
      k)}}_{k=\kdiam}  = \alphasdiam{2} + \alphasdiam{3} + \alphasdiam{4} + \order{\eps{}^5}.
\end{equation}
The leading-order term is second order in the slow-roll parameters and reads
\begin{equation}
\alphasdiam{2} = -2\epsdiam{1} \epsdiam{2} - \epsdiam{2}
\epsdiam{3} +\deldiam{1} \deldiam{2},
\end{equation}
while the third order term is given by
\begin{equation}
  \begin{aligned}
\alphasdiam{3} & = -6 \epsdiam{1}^2 \epsdiam{2} - \left(2C+3\right)
\epsdiam{1}\epsdiam{2}^{2} - C \left(\epsdiam{2} \epsdiam{3}^{2} +
\epsdiam{2} \epsdiam{3} \epsdiam{4} \right) -\left(2C+4\right)
\epsdiam{1} \epsdiam{2} \epsdiam{3} + 5 \epsdiam{1} \epsdiam{2}
\deldiam{1} + 2 \epsdiam{2} \epsdiam{3} \deldiam{1} \\ & + 4
\epsdiam{1} \deldiam{1} \deldiam{2} + \epsdiam{2} \deldiam{1}
\deldiam{2} - 3\deldiam{1}^{2} \deldiam{2} + \left(C + 2\right)\left(
\deldiam{1} \deldiam{2}^{2} + \deldiam{1} \deldiam{2}
\deldiam{3}\right).
\end{aligned}
\end{equation}
Both $\alphasdiam{2}$ and $\alphasdiam{3}$ match the expressions
previously derived in Ref.~\cite{Jimenez:2013xwa}. The new term is the
fourth order and reads
\begin{equation}
  \begin{aligned}
\alphasdiam{4} & = -12 \epsdiam{1}^3 \epsdiam{2} + \left(2 \pi^{2} -
14 C - 35\right)\epsdiam{1}^2 \epsdiam{2}^2 + \left(\pi^{2} - 8 C-
19\right) \epsdiam{1}^2\epsdiam{2}\epsdiam{3} + \frac{1}{12} {\left(7
  \pi^{2} - 12 C^{2} - 36 C - 84\right)} \epsdiam{1} \epsdiam{2}^{3}
\\ & + \left(\frac{7}{4} \pi^{2} - 3 C^{2} - 10 C -21\right)
\epsdiam{1}\epsdiam{2}^2\epsdiam{3} + \left(\frac{\pi^2}{8} -
\frac{3}{2} C^{2}\right)\epsdiam{2}\epsdiam{3}^2\epsdiam{4} +
\left(\frac{\pi^2}{4} - 2\right)\left( 2 \epsdiam{2}^2 \epsdiam{3}^2 +
\epsdiam{2}^2 \epsdiam{3} \epsdiam{4}\right) \\ & + \left(\frac{7}{12}
\pi^{2} - C^{2} - 5 C - 6\right)\left(\epsdiam{1} \epsdiam{2}
\epsdiam{3}^2 + \epsdiam{1}\epsdiam{2}\epsdiam{3}\epsdiam{4} \right) +
\frac{1}{24} {\left(\pi^{2} - 12 C^{2}\right)} \left( \epsdiam{2}
\epsdiam{3}^{3} + \epsdiam{2} \epsdiam{3} \epsdiam{4}^{2} +
\epsdiam{2} \epsdiam{3} \epsdiam{4} \epsdiam{5} \right) \\ & - 15
\epsdiam{1}\deldiam{1}^2\deldiam{2} - \left(\frac{\pi^2}{2} - 5 C -
12\right)\left( \epsdiam{1}\deldiam{1}\deldiam{2}^2 +
\epsdiam{1}\deldiam{1}\deldiam{2}\deldiam{3}\right) -
\left(\frac{\pi^2}{4} - C - 3\right)\left(
\epsdiam{2}\deldiam{1}\deldiam{2}^2 +
\epsdiam{2}\deldiam{1}\deldiam{2}\deldiam{3} \right) \\ & - 3
\epsdiam{2}\deldiam{1}^2\deldiam{2} + 9 \epsdiam{1}^{2}\deldiam{1}
\deldiam{2} - 9 \epsdiam{1}\epsdiam{2}\deldiam{1}^2 - 3
\epsdiam{1}\epsdiam{3}\deldiam{1}^2 - \left(\pi^{2} - 9 C -
25\right)\epsdiam{1}\epsdiam{2}\deldiam{1}\deldiam{2} -
\left(\frac{\pi^2}{2} - 3 C -6\right)
\epsdiam{2}\epsdiam{3}\deldiam{1}\deldiam{2} \\ & + 21
\epsdiam{1}^2\epsdiam{2}\deldiam{1} - \left(\frac{\pi^2}{2} - 7 C -
16\right)\epsdiam{1} \epsdiam{2}^2 \deldiam{1} - \left(\frac{\pi^2}{4}
- 3 C - 3\right)\epsdiam{2} \epsdiam{3}^2 \deldiam{1} -
\left(\frac{\pi^2}{2} - 7 C - 19\right)
\epsdiam{1}\epsdiam{2}\epsdiam{3}\deldiam{1} \\ & -
\left(\frac{\pi^2}{4} - 3 C - 3\right) \epsdiam{2} \epsdiam{3}
\epsdiam{4}\deldiam{1} + 6 \deldiam{1}^{3} \deldiam{2} + \frac{1}{2}
\left(\pi^{2} - 14 C - 36\right) \deldiam{1}^{2} \deldiam{2}^{2} +
\left(\frac{\pi^2}{4} - 4 C - 10 \right)
\deldiam{1}^2\deldiam{2}\deldiam{3} \\ & - \left(\frac{\pi^{2}}{8} -
\frac{3}{2} C^{2} - 6 C - 6\right)\deldiam{1}\deldiam{2}^2\deldiam{3}
- \frac{1}{24} {\left(\pi^{2} - 12 C^{2} - 48 C - 48\right)} \left(
\deldiam{1} \deldiam{2}^{3} + \deldiam{1} \deldiam{2} \deldiam{3}^{2}
+ \deldiam{1} \deldiam{2} \deldiam{3} \deldiam{4} \right).
  \end{aligned}
\end{equation}

Next is the running of the running of the scalar spectral index
\begin{equation}
\betaS \equiv \eval{\dv[3]{\ln\left[\calPz(k)\right]}{(\ln
    k)}}_{k=\kdiam}=\betasdiam{3} + \betasdiam{4} + \order{\eps{}^5},
\end{equation}
with
\begin{equation}
\betasdiam{3} =- 2 \epsdiam{1} \epsdiam{2}^{2} - 2
\epsdiam{1}\epsdiam{2} \epsdiam{3} - \epsdiam{2} \epsdiam{3}^{2} -
\epsdiam{2}\epsdiam{3} \epsdiam{4} + \deldiam{1} \deldiam{2}^{2} +
\deldiam{1} \deldiam{2} \deldiam{3},
\end{equation}
and
\begin{equation}
  \begin{aligned}
\betasdiam{4} & = -14 \epsdiam{1}^{2}\epsdiam{2}^{2} - 8
\epsdiam{1}^2\epsdiam{2}\epsdiam{3} - \left(2 C + 3\right) \epsdiam{1}
\epsdiam{2}^{3} - C \left(\epsdiam{2} \epsdiam{3}^{3} + 3
\epsdiam{2}\epsdiam{3}^2\epsdiam{4} + \epsdiam{2} \epsdiam{3}
\epsdiam{4}^{2} + \epsdiam{2} \epsdiam{3} \epsdiam{4}
\epsdiam{5}\right) \\ & - \left(2 C +
5\right)\left(\epsdiam{1}\epsdiam{2}
\epsdiam{3}^{2}+\epsdiam{1}\epsdiam{2}\epsdiam{3}\epsdiam{4} \right)
-\left(6C+10\right) \epsdiam{1}\epsdiam{2}^2\epsdiam{3} + 5
\epsdiam{1} \deldiam{1} \deldiam{2}^{2} + 5 \epsdiam{1} \deldiam{1}
\deldiam{2} \deldiam{3} + \epsdiam{2}\deldiam{1} \deldiam{2}^{2} +
\epsdiam{2}\deldiam{1} \deldiam{2} \deldiam{3} \\ & + 9 \epsdiam{1}
\epsdiam{2}\deldiam{1} \deldiam{2} + 3
\epsdiam{2}\epsdiam{3}\deldiam{1}\deldiam{2} + 3 \epsdiam{2}
\epsdiam{3}^{2}\deldiam{1} + 7 \epsdiam{1}\epsdiam{2}^{2}\deldiam{1} +
7 \epsdiam{1}\epsdiam{2}\epsdiam{3}\deldiam{1} + 3
\epsdiam{2}\epsdiam{3}\epsdiam{4}\deldiam{1} - 7 \deldiam{1}^{2}
\deldiam{2}^{2} - 4 \deldiam{1}^{2} \deldiam{2}\deldiam{3} \\ & +
\left(C + 2\right) \left( \deldiam{1} \deldiam{2}^{3} + \deldiam{1}
\deldiam{2} \deldiam{3}^{2} \deldiam{1} \deldiam{2}
\deldiam{3}\deldiam{4} + 3 \deldiam{1} \deldiam{2}^{2}\deldiam{3}
\right).
  \end{aligned}
\end{equation}
Finally, we also get the leading-order term of the running of the
running of the running of the scalar spectral index
\begin{equation}
\gammaS \equiv \eval{\dv[4]{\ln\left[\calPz(k)\right]}{(\ln
    k)}}_{k=\kdiam} = \gammasdiam{4} + \order{\eps{}^5},
\end{equation}
where
\begin{equation}
\begin{aligned}
  \gammasdiam{4} &= - 2 \epsdiam{1} \epsdiam{2}^{3} - 6 \epsdiam{1}
  \epsdiam{2}^{2} \epsdiam{3} - 2 \epsdiam{1} \epsdiam{2}
  \epsdiam{3}^{2} - \epsdiam{2} \epsdiam{3}^{3} - \epsdiam{2}
  \epsdiam{3} \epsdiam{4}^{2} - \epsdiam{2} \epsdiam{3} \epsdiam{4}
  \epsdiam{5} - 2 \epsdiam{1} \epsdiam{2} \epsdiam{3} \epsdiam{4} - 3
  \epsdiam{2} \epsdiam{3}^{2} \epsdiam{4} \\ & + \deldiam{1}
  \deldiam{2}^{3} + 3 \deldiam{1} \deldiam{2}^{2} \deldiam{3} +
  \deldiam{1} \deldiam{2} \deldiam{3}^{2} + \deldiam{1} \deldiam{2}
  \deldiam{3} \deldiam{4}.
\end{aligned}
\end{equation}

\subsection{Tensor power-law parameters}

The equivalent power-law quantities can be derived in the exact same way for
the tensor modes. From \cref{eq:calPhsr3ast}, implementing the change
of pivot of \cref{eq:ast2diam}, the tensor spectral index is defined
by
\begin{equation}
\nT \equiv \eval{\dv{\ln\left[\calPh(k)\right]}{\ln k}}_{k=\kdiam} =
\ntdiam{1} + \ntdiam{2} + \ntdiam{3} + \ntdiam{4} + \order{\eps{}^5}.
\end{equation}
We have
\begin{equation}
  \begin{aligned}
    \ntdiam{1} & = - 2\epsdiam{1}, \\
    \ntdiam{2} & = - 2\epsdiam{1}^2 - 2\left(C+1 - \ln\csdiam\right)
    \epsdiam{1}\epsdiam{2},\\
    \ntdiam{3} & = - 2  \epsdiam{1}^{3} + \left(\pi^{2} - 6 C  - 14 + 6 \ln^2\csdiam\right)
    \epsdiam{1}^{2}\epsdiam{2} + \left[\frac{\pi^2}{12}  - C^{2} + 2
      \left(C + 1\right) \ln\csdiam - \ln^2\csdiam
       - 2 C - 2\right] \left( \epsdiam{1}
    \epsdiam{2}^{2} +  \epsdiam{1} \epsdiam{2} \epsdiam{3} \right),
  \end{aligned}
\end{equation}
and the N4LO correction
\begin{equation}
  \begin{aligned}
    \ntdiam{4} &= - 2 \epsdiam{1}^{4} + \left[4 \pi^{2} - 12 C +
    12 \ln\csdiam - 14 \zeta(3) -  28\right]\epsdiam{1}^3 \epsdiam{2}
    \\ & + \left[\frac{31}{12} \pi^{2} + 2 \left(\pi^{2} - 16\right) C - 7 C^{2} - 2 \left(\pi^{2} - 7 C -
        16\right) \ln\csdiam - 7 \ln^{2}\csdiam - 30\right]
    \epsdiam{1}^2 \epsdiam{2}^2
    \\ & + \left[-\frac{C^3}{3} - \left(C + 1\right) \ln^{2}\csdiam +
      \frac{1}{3}\ln^3\csdiam
      + \frac{\pi^2}{12} + \frac{1}{12} \left(\pi^{2} - 24\right) C -
      C^{2} - \frac{1}{12} \left(\pi^{2} - 12  C^{2}
      - 24 C - 24\right) \ln\csdiam \right. \\ & \left. -
      \frac{2}{3}\zeta(3) - \frac{2}{3}\right]
    \left(\epsdiam{1}\epsdiam{2}^3 + 3
    \epsdiam{1}\epsdiam{2}^2\epsdiam{3} +\epsdiam{1}
    \epsdiam{2} \epsdiam{3}^2 + \epsdiam{1}\epsdiam{2}\epsdiam{3}\epsdiam{4} \right)
    \\ & + \left[\frac{4}{3} \pi^{2} + \left(\pi^{2} - 16\right)
      C - 4 C^{2} - \left(\pi^{2} - 8 C - 16\right) \ln\csdiam - 4 \ln^2\csdiam - 16\right]
    \epsdiam{1}^2 \epsdiam{2} \epsdiam{3}.
  \end{aligned}
\end{equation}

The running of the tensor spectral index is defined by
\begin{equation}
  \alphaT  \equiv \eval{\dv[2]{\ln\left[\calPh(k)\right]}{(\ln
      k)}}_{k=\kdiam}  = \alphatdiam{2} + \alphatdiam{3} + \alphatdiam{4} + \order{\eps{}^5},
\end{equation}
and we find
\begin{equation}
  \begin{aligned}
    \alphatdiam{2} & = - 2 \epsdiam{1} \epsdiam{2}, \\
    \alphatdiam{3} & = - 6 \epsdiam{1}^2 \epsdiam{2} -
    2\left(C+1-\ln\csdiam\right) \left(\epsdiam{1}\epsdiam{2}^2 + \epsdiam{1}\epsdiam{2}\epsdiam{3}\right),
  \end{aligned}
\end{equation}
and
\begin{equation}
  \begin{aligned}
    \alphatdiam{4} & = - 12 \epsdiam{1}^{3} \epsdiam{2} +
    \left(2\pi^{2} - 14 C + 14 \ln\csdiam - 32\right)
    \epsdiam{1}^2 \epsdiam{2}^2 + \left(\pi^{2} - 8 C + 8 \ln\csdiam - 16\right)
    \epsdiam{1}^2 \epsdiam{2} \epsdiam{3}
      \\ & + \left[\frac{\pi^2}{12}  -
      C^{2} + 2 \left(C + 1\right) \ln\csdiam - \ln^2\csdiam - 2 C - 2\right]\left( \epsdiam{1}
              \epsdiam{2}^{3} + 3  \epsdiam{1} \epsdiam{2}^{2}
              \epsdiam{3} + \epsdiam{1} \epsdiam{2} \epsdiam{3}^{2} +
              \epsdiam{1} \epsdiam{2} \epsdiam{3} \epsdiam{4} \right).
  \end{aligned}
\end{equation}

The running of the running of the tensor spectral index is
\begin{equation}
\betaT \equiv \eval{\dv[3]{\ln\left[\calPh(k)\right]}{(\ln
    k)}}_{k=\kdiam}=\betatdiam{3} + \betatdiam{4} + \order{\eps{}^5},
\end{equation}
with
\begin{equation}
  \begin{aligned}
    \betatdiam{3} & = - 2 \epsdiam{1} \epsdiam{2}^2 - 2 \epsdiam{1}
    \epsdiam{2} \epsdiam{3},\\
    \betatdiam{4} & = -14 \epsdiam{1}^2 \epsdiam{2}^2 - 8
    \epsdiam{1}^2 \epsdiam{2} \epsdiam{3}
    - 2\left(C +1 - \ln\csdiam\right)\left( \epsdiam{1}
              \epsdiam{2}^{3} + 3  \epsdiam{1} \epsdiam{2}^{2}
              \epsdiam{3} + \epsdiam{1} \epsdiam{2} \epsdiam{3}^{2} +
              \epsdiam{1} \epsdiam{2} \epsdiam{3} \epsdiam{4} \right).
  \end{aligned}
\end{equation}

Finally, the running of the running of the running of the tensor
spectral index is defined by
\begin{equation}
\gammaT \equiv \eval{\dv[4]{\ln\left[\calPh(k)\right]}{(\ln
    k)}}_{k=\kdiam} = \gammatdiam{4} + \order{\eps{}^5},
\end{equation}
and the N4LO correction reads
\begin{equation}
\gammatdiam{4} = -2 \left( \epsdiam{1}
              \epsdiam{2}^{3} + 3  \epsdiam{1} \epsdiam{2}^{2}
              \epsdiam{3} + \epsdiam{1} \epsdiam{2} \epsdiam{3}^{2} +
              \epsdiam{1} \epsdiam{2} \epsdiam{3} \epsdiam{4} \right).
\end{equation}

\subsection{Tensor-to-scalar ratio}

Our calculation allows up to derive an expression for the
tensor-to-scalar ratio $r$ at fourth order in the slow-roll
parameters. From the definition
\begin{equation}
r \equiv \dfrac{\calPh(\kdiam)}{\calPz(\kdiam)}\,,
\end{equation}
using \cref{eq:calPhsr3ast,eq:calPzsr3diam,eq:ast2diam}, we obtain
\begin{equation}
  \begin{aligned}
r & = 16 \epsdiam{1} \csdiam \left[1+ \rdiam{1} + \rdiam{2} + \rdiam{3}
\right] + \order{\eps{}^5},
  \end{aligned}
\end{equation}
and we recover the already known terms, namely
\begin{equation}
  \rdiam{1}  = C \epsdiam{2} - \left(C+2\right)\deldiam{1} + 2 \epsdiam{1} \ln\csdiam ,\\
\end{equation}
and
\begin{equation}
\begin{aligned}
  \rdiam{2} & = - \frac{1}{8} \left(\pi^{2} - 4 C^{2} - 8\right)
  \epsdiam{2}^{2} - \left[\frac{\pi^2}{2} - 2 \left(2 C + 1\right)
    \ln\csdiam + \ln^2\csdiam - C - 4\right] \epsdiam{1}\epsdiam{2} -
  \frac{1}{24} {\left(\pi^{2} - 12 C^{2}\right)} \epsdiam{2}
  \epsdiam{3} \\ & + 2 \ln\csdiam \left(\ln\csdiam + 1
  \right)\epsdiam{1}^2 +\left[\frac{\pi^2}{2} - 2 \left(C +
    2\right)\ln\csdiam - 3 C - 8\right] \epsdiam{1} \deldiam{1} +
  \left(\frac{\pi^2}{4} - C^{2} - 3 C - 3\right)\epsdiam{2}\deldiam{1}
  \\ & -\frac{1}{8} \left(\pi^{2} - 4 C^{2} - 24 C - 40\right)
  \deldiam{1}^{2} + \frac{1}{24}\left(\pi^{2} - 12 C^{2} - 48 C -
  48\right) \deldiam{1} \deldiam{2}.
\end{aligned}
\end{equation}
The new correction reads
\begin{equation}
  \begin{aligned}
    \rdiam{3} & = \frac{1}{24}  \left[4  C^{3} - 3  \left(\pi^{2} -
   8\right) C + 14  \zeta(3) - 16\right] \epsdiam{2}^{3}  -
    \frac{1}{24}  \left[\pi^{2} C - 4  C^{3} - 8  \zeta(3) + 16\right]
\left(\epsdiam{2} \epsdiam{3}^{2} + \epsdiam{2} \epsdiam{3}
\epsdiam{4} \right) \\ &+ \left[\left(6 C + 1\right)\ln^2\csdiam - 2 \ln^3\csdiam -
    \pi^{2} - 2 \left(\pi^{2} - 5 C - 11\right) \ln\csdiam + C +
    7\zeta(3)\right] \epsdiam{1}^2 \epsdiam{2}
\\ & + \left[\frac{1}{2} C^{3} - \frac{1}{24} \left(7  \pi^{2} - 48\right)
  C\right] \epsdiam{2}^2 \epsdiam{3} + \left(\frac{4}{3}
      \ln^3\csdiam + 4 \ln^2\csdiam + 2\ln\csdiam \right)\epsdiam{1}^3
    \\& - \left[\left(C + 1\right)\ln^2\csdiam - \frac{1}{3}\ln^3\csdiam +
  \frac{\pi^2}{12} + \frac{C}{2} \left(\pi^{2} - 8\right) - C^{2} +
  \frac{1}{6}\left(\pi^{2} - 12 C^{2} - 12 C - 12\right)\ln\csdiam
  \right] \epsdiam{1} \epsdiam{2} \epsdiam{3}
    \\&
     + \left[\frac{\ln^3\csdiam}{3} -\left(2 C + 1\right)\ln^2\csdiam
     - \frac{19}{24} \pi^{2} - \left(\pi^{2} -
    9\right) C + \frac{3}{2} C^{2} - \frac{1}{3}\left(\pi^{2} - 12
    C^{2} - 12 C - 12\right)\ln\csdiam + \frac{7}{2} \zeta(3) + 2
    \right]\epsdiam{1}\epsdiam{2}^2 \\ &
+ \left[-\frac{9}{4} \pi^{2} - \frac{1}{2} \left(\pi^{2} -
      36\right) C + 3 C^{2} - \frac{1}{4} \left(\pi^{2} - 4 C^{2} - 24
      C - 40\right)\ln\csdiam + \frac{7}{2} \zeta(3) + 29\right]
  \epsdiam{1}\deldiam{1}^2
  \\ &
  + \left[\frac{1}{2} C^{3} - \pi^{2} - \frac{1}{8} \left(3\pi^{2} - 88\right) C + 4 C^{2} + \frac{7}{4} \zeta(3) +
    \frac{19}{2}\right]\epsdiam{2}\deldiam{1}^2
\\&+ \left[\frac{7}{6}\pi^{2} + \frac{1}{2} \left(\pi^{2} - 20\right)
    C - 2  C^{2} + \frac{1}{12}\left(\pi^{2} - 12 C^{2} - 48 C -
    48\right) \ln\csdiam - 14\right] \epsdiam{1}\deldiam{1}\deldiam{2}
  \\ & +
   \left[-\frac{1}{2} C^{3} + \frac{13}{24} \pi^{2} + \frac{1}{24} \left(7 \pi^{2} - 120\right) C - \frac{5}{2}
     C^{2} - 5\right]\epsdiam{2}\deldiam{1}\deldiam{2}
\\&   + \left[2 \pi^{2} -2 \left(C + 2\right)\ln^2\csdiam  + \left(\pi^{2} -
  8 C - 20\right)\ln\csdiam - 5 C - 7 \zeta(3) - 14 \right]
\epsdiam{1}^{2} \deldiam{1}
\\&  + \left[ \frac{\pi^2}{2} -\frac{1}{2} C^{3}  +
    \frac{1}{8} \left(3 \, \pi^{2} - 32\right) C - 2 C^{2} -
    \frac{7}{4}  \zeta(3) - 2\right]\epsdiam{2}^2 \deldiam{1}
\\ & + \left[\left(C + 2\right)\ln^2\csdiam + \frac{77}{24}\pi^{2} +
    \frac{3}{2}\left(\pi^{2} - 18\right) C - \frac{13}{2} C^{2} +
    \frac{1}{2} \left(\pi^{2} - 8 C^{2}- 24 C - 20\right)
         \ln\csdiam - 7 \zeta(3) - 24\right]
  \epsdiam{1}\epsdiam{2}\deldiam{1}
       \\ & + \left[\frac{\pi^2}{6} -\frac{1}{2} C^{3} + \frac{1}{24}
      \left(7 \pi^{2} - 72\right) C - 2  C^{2}\right] \epsdiam{2}\epsdiam{3}\deldiam{1}
      -\frac{1}{24}  \left[4  C^{3} - 12  \pi^{2} - 3
  {\left(\pi^{2} - 64\right)} C + 48  C^{2} + 14  \zeta(3) +
  260\right] \deldiam{1}^{3} \\& + \frac{1}{24} \left[12  C^{3} - 17
  \pi^{2} - 7  {\left(\pi^{2} - 48\right)} C + 108  C^{2} +
  360\right] \deldiam{1}^{2} \deldiam{2} \\ &
- \frac{1}{24}  \left[4  C^{3} - 2  \pi^{2} - \left(\pi^{2} -
    48\right) C + 24  C^{2} + 8  \zeta(3) + 32\right] \left(\deldiam{1}
\deldiam{2}^{2} + \deldiam{1} \deldiam{2} \deldiam{3}\right).
  \end{aligned}
\end{equation}

\end{widetext}
\bibliography{biblio}

\end{document}